\documentclass[usenatbib]{mn2e}

\usepackage{txfonts}
\usepackage{graphicx}

\newcommand{\hess}{{H.E.S.S.}}
\newcommand{\fer}{{\sl {\it Fermi}}}
\newcommand{\fla}{\fer-LAT}
\newcommand{\chandra}{{\sl Chandra}}
\newcommand{\gr}{$\gamma$-ray}

\newcommand{\mec}{{ m}_{ e}{ c}^2}
\newcommand{\ergcms}{\ensuremath{\mathrm{erg}\ \mathrm{cm}^{-2}\ \mathrm{s}^{-1}}}%
\newcommand{\g}{\gamma}

\newcommand\arcdeg{\mbox{$^\circ$}}%
\newcommand{\hmsh}{\mbox{$^{\mathrm h}$}}%
\newcommand{\hmsm}{\mbox{$^{\mathrm m}$}}%
\newcommand{\hmss}{\mbox{$^{\mathrm s}$}}%
\newcommand{\HMS}[3]{$#1\hmsh\,#2\hmsm\,#3\hmss$}
\newcommand{\DMS}[3]{$#1\arcdeg\,#2\arcmin\,#3\arcsec$}

\newcommand{\erg}{\mathrm{erg}}

\newcommand{\s}{\mathrm{s}}

\newcommand{\Gauss}{\mathrm{G}}

\usepackage{lineno}
\modulolinenumbers[5]

\begin{document}

\title{From Radio to TeV: The surprising Spectral Energy Distribution of AP~Librae}

\author[D.~A.~Sanchez et al.]{
D.A.~Sanchez,$^1$\thanks{david.sanchez@lapp.in2p3.fr} B.~Giebels,$^2$ P.~Fortin,$^3$ D.~Horan,$^2$  A.~Szostek,$^4$, S. Fegan,$^2$\newauthor A.-K.~Baczko,$^{5,6}$ J.~Finke,$^7$  M.L.~Kadler,$^6$ Y.Y.~Kovalev,$^{8,9}$ M.L.~Lister$^{10}$ A.B.~Pushkarev$^{11,12,9}$\newauthor T.~Savolainen$^{13,9}$\\
$^1$Laboratoire d'Annecy-le-Vieux de Physique des Particules, Universit\'{e} de Savoie, CNRS/IN2P3, F-74941 Annecy-le-Vieux, France\\
$^2$Laboratoire Leprince-Ringuet, Ecole Polytechnique, CNRS/IN2P3, F-91128 Palaiseau, France \\
$^3$Fred Lawrence Whipple Observatory, Harvard-Smithsonian Center for Astrophysics, Amado, AZ 85645, USA\\
$^4$ Kavli Institute for Particle Astrophysics and Cosmology, Department of Physics and SLAC National Accelerator Laboratory, Stanford University, Stanford, CA
94305, USA
$^5$Dr Karl Remeis-Observatory  ECAP, Astronomical Institute, Friedrich-Alexander University Erlangen-Nuremberg, Sternwartstr. 7, 96049 Bamberg, Germany\\
$^6$Lehrstuhl f\"ur Astronomie, Universit\"at W\"urzburg, Campus Hubland Nord, Emil-Fischer-Strasse 31, 97074, W\"urzburg, Germany\\  
$^7$U.S.\ Naval Research Laboratory, Code 7653, 4555 Overlook Ave. SW, Washington, DC, 20375-5352\\
$^8$Astro Space Center of Lebedev Physical Institute, Profsoyuznaya 84/32, 117997 Moscow, Russia\\
$^9$Max-Planck-Institut f\"ur Radioastronomie, Auf dem H\"ugel 69, 53121 Bonn, Germany\\
$^{10}$Department of Physics, Purdue University, West Lafayette, IN 47906, USA\\
$^{11}$Crimean Astrophysical Observatory, 98409 Nauchny, Crimea, Russia\\
$^{12}$Pulkovo Observatory, Pulkovskoe Chaussee 65/1, 196140 St. Petersburg, Russia \\
$^{13}$Aalto University Mets\"ahovi Radio Observatory, Mets\"ahovintie 114, 02540 Kylm\"al\"a, Finland
}


\maketitle

\begin{abstract}
Following the discovery of high-energy (HE; $E>10\,{\rm MeV}$) and very-high-energy (VHE; $E>100\,{\rm GeV}$)
$\gamma$-ray emission from the low-frequency-peaked BL~Lac (LBL) object AP Librae, its electromagnetic
spectrum is studied over 60 octaves in energy. Contemporaneous data in radio, optical and UV together with the (non simultaneous)
$\gamma$-ray data are used to construct the most precise spectral energy distribution of this source. The data
have been found to be modeled with difficulties with single zone homogeneous leptonic synchrotron self-Compton
(SSC) radiative scenarios due to the unprecedented width of the high-energy component when compared to the
lower-energy component. The two other LBL objects also detected at VHE appear to have similar modeling
difficulties.  Nevertheless, VHE $\gamma$ rays produced in the extended jet could account for the VHE flux
observed by H.E.S.S.

\end{abstract}


\maketitle

\section{Introduction}

Blazars are among the most energetic objects in the Universe that exhibit
non-thermal electromagnetic spectra from radio up to very-high-energy (VHE,
E$>$100 GeV) $\gamma$-rays, with a two-component spectral energy distribution
(SED) structure in a $\nu f(\nu)$ representation. Multi-wavelength data are of paramount importance to understand the mechanisms at play in the jet.


Blazars are divided into two classes: flat spectrum radio quasars (FSRQs) and
BL~Lacertae (BL~Lac) objects, the latter being sub-divided into high-frequency-peaked BL~Lac (HBL) and low-frequency-peaked BL~Lac (LBL). The distinction
between HBL and LBL classes is based on the low-energy peak position
\citep{1995ApJ...444..567P}. HBL objects present a peak in the UV or X-ray range
while the peak of LBL objects is located at lower energies (i.e., in optical
wavelengths).

So far, the vast majority of BL Lac objects detected in VHE belong to the HBL
sub-class\footnote{To keep track of the number of detected object, an up-to-date
VHE $\gamma$-ray catalogue can be found in the TeVCat {\tt
http://tevcat.uchicago.edu}}. The SEDs of HBL objects are often successfully
modeled with a synchrotron self-Compton (SSC) model, in which the low-energy
emission is produced by synchrotron radiation of relativistic electrons, and the
high-energy component by inverse Compton scattering off the same synchrotron
photons. HBL are the dominant class of extragalactic objects detected by
ground-based Atmospheric \v{C}erenkov Telescopes (ACTs) in the TeV $\gamma$-ray
regime.

Only a few TeV emitters belong to the LBL sub-class and, among them, AP~Librae
\citep[$z=0.049$,][]{2009MNRAS.399..683J} was recently detected by the H.E.S.S. collaboration
\citep{ApLibHESS} with a flux of $8.78 \pm 1.54_{\rm stat} \pm 1.76_{\rm sys} \times 10^{-12} \,{\rm cm}^{-2}
      {\rm s}^{-1}$ above 130 GeV and a photon index $\Gamma = 2.65\pm0.19_{\rm stat}\pm0.20_{\rm sys}$
      matching well the spectrum measured by the \fer\ Large Area telescope (LAT) in the high energy (HE,
      100\,MeV$<$E$<$300\,GeV) range. Remarkably, the spectral break between the HE and VHE ranges is the
      smallest ever measured for an LBL object but cannot be explained by extragalactic background light (EBL)
      attenuation only \citep{2013arXiv1303.5923S}. In this work, VHE and HE data have been extracted from \citet{ApLibHESS}

After the announcement of this detection by the \hess\ collaboration \citep{2010ATel.2743....1H}, {\it Swift}
and {\it RXTE} data were taken creating contemporaneous spectra in X-ray and UV bands. Analysis and
results are presented in sections \S\ref{sub:swift} and \S\ref{UV_OPTIC}. Archival observation by
\chandra\ (\S\ref{sub:chandra}) have been analyzed in this work, revealing the first X-ray extended jet for a
VHE blazar.  At longer wavelengths, AP~Librae is one of the targets of different monitoring programs such as
SMARTS (\S\ref{UV_OPTIC}) and the MOJAVE program (\S\ref{mojave}) which provide long-term optical and VLBA
measurements. The VHE detection, together with lower energy-data presented in this paper, enable to draw the
first complete SED of this source and to probe mechanisms at play in LBL objects. The broadband SED is then
discussed in the framework of different emission models in \S\ref{discussion} .

Throughout this paper a $\Lambda$CDM cosmology with H$_0 = 71$\,km\,s$^{-1}$\,Mpc$^{-1}$, $\Omega_\Lambda =
0.73$ and $\Omega_{\rm M}\ =\ 0.27$ is assumed, resulting in a luminosity distance for AP~Librae of D$_{\rm L}
= 215$ Mpc and a linear scale of 0.947\,kpc per arcsecond \citep{2006PASP..118.1711W}.

\section{Multi-wavelength Observations}
\label{sec:observations}

\subsection{{\it Swift}-XRT and {\it RXTE}-PCA observations}
\label{sub:swift}


X-ray observations of AP~Librae during the period of interest were retrieved using the HEASARC archive. Four
consecutive daily observations (ObsID 95141) of $\simeq 3\,{\rm ks}$ each were carried out between 10--14 July
2010 with {\sl RXTE} \citep{1996SPIE.2808...59J}, with a total exposure of $\simeq 13\,{\rm ks}$. The
STANDARD2 {\sl RXTE-}PCA data were extracted using the ftools in the HEASOFT 6.16 software package provided by
NASA/GSFC and filtered using the RXTE Guest Observer Facility recommended criteria. Only signals from the top
layer (X1L and X1R) of Proportional Counter Unit 2 (PCU2) were used to extract spectra in the $3-50\,{\rm
  keV}$ range, using the faint-background model. The obtained daily light curve has an average rate of
$0.44\,{\rm counts}\,{\rm s^{-1}}$, a variance of $0.03 \,{\rm counts}\,{\rm s^{-2}} $ compatible with its
expected variance of $0.02 \,{\rm counts}\,{\rm s^{-2}} $ if the source were constant, and a chi-square
probability of constancy of $27\%$, hence no variability is present over the span of 4 days.

During the period of interest, seven observations were carried out by the {\it Swift} mission
\citep{2005SSRv..120..165B}, between 20 February 2010 and 16 August 2011 (ObsID 36341005 to 36341011), of
which one $5\,{\rm ks}$ observation was carried out on 7 July 2010, near the {\sl RXTE} observation. However,
the short observation in ObsID 36341009 was skipped. The photon-counting (PC) mode data are processed with the
standard {\tt xrtpipeline} tool (HEASOFT 6.16), with the source and background-extraction regions defined as a
20-pixel (∼4.7 arcsec) and a 40-pixel radius circle respectively, the latter being centered nearby the former
without overlapping. All exposures show a source with a stable average count rate of $\simeq 0.12\,{\rm
  counts}\,{\rm s^{-1}}$. Also the large $5\,{\rm ks}$ XRT-PC light curve shows the source with an average
count rate of $(0.13\pm0.02)\,{\rm s^{-1}}$ and an r.m.s. of $\simeq 0.01\,{\rm s^{-1}}$ for which no
variability could be found with a $99\%$ confidence level upper limit on the fractional variance (as defined
in \citealt{Vaughan}) $F_{\rm var}$ of 0.95. Using this count rate in WebPIMMS from {\tt HEASARC}, a RXTE-PCA
count rate of $\simeq 0.6\,{\rm counts}\,{\rm s^{-1}}$ is predicted, compatible with the value actually
observed of $0.44\,{\rm counts}\,{\rm s^{-1}}$ hinting at the fact that the source was probably in the same state during observations of both
observatories. Given the low count rate, no pile-up is expected in PC mode, which is confirmed by the
acceptable fit of a King profile to the PSF of all observations.

\begin{figure*}
\includegraphics[width=0.99 \textwidth]{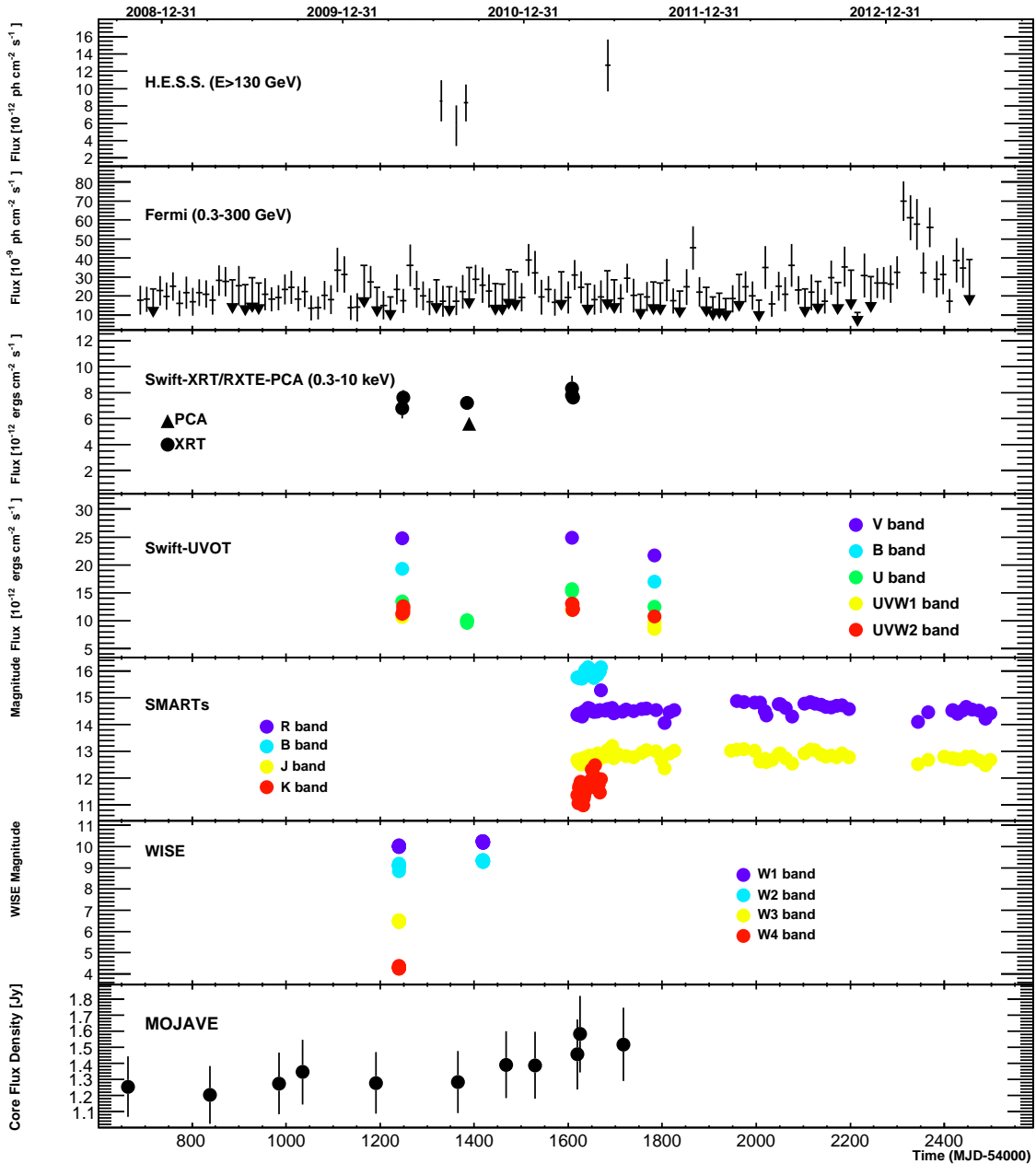}
\caption{Light curves of AP~Librae in, from top to bottom, VHE, HE, X-rays, UV, Optical and radio (15 GHz) wavebands. The 4 RXTE observations (ObsID 95141) were merged together and the seven {\it
Swift} observations (ObsID 36341005 to 36341011) are shown individually.}
\label{FIG:LC}	
\end{figure*}

Spectral fitting of all ObsIDs was performed with PyXspec v1.0.4 \citep{1996ASPC..101...17A}, using a response
matrix for the combined PCA data set generated by the ftool {\tt pcarsp v11.7.1}, and dedicated Ancillary
Response Functions (ARFs) for all XRT data sets generated by {\tt xrtmkarf} (along with the latest spectral
redistribution matrices {\tt swxpc0to12s6\_20110101v014} from CALDB). Spectra from all ObsIDs were rebinned to
have at least 20 counts per bin using {\tt grppha}, channels 0 to 29 were ignored in the XRT-PC data, and only
the 3--50$\,{\rm keV}$ range is used in the PCA data. All data sets are fit to a power-law model ${\rm d}N/{\rm
  d}E=N_0(E/E_0)^{-\Gamma_{\rm X}}$, where $N_0$ is the normalization factor at a chosen reference energy
$E_0=1$~keV and $\Gamma_{\rm X}$ the photon index. Using the Leiden/Argentine/Bonn (LAB) Survey of Galactic HI
\citep{2005A&A...440..775K} weighted average hydrogen column density of $N_{\rm H}=8.14\times10^{20}\,{\rm
  cm^{-2}}$, good fits are obtained for the power-law function ($P(\chi^2)=0.18-0.9$) with a photon index of
$\Gamma_{\rm X}\simeq1.55$ on average. All XRT observations were also summed, a new exposure file built with
      {\tt ximage}, and a new ARF for the summed spectrum. This latter spectrum extends up to $\simeq 7\,{\rm
        keV}$. Another spectrum was derived this time limited to 1 count/bin, to allow an extension to higher
      energies, and was fitted using {\tt statistic cstat} required in the case of Poisson data. The fit
      parameters are entirely compatible with those obtained using $\chi^2$ statistics, but the spectrum
      extends up to $\simeq 10\,{\rm keV}$. All fit parameters, along with the unabsorbed 0.3--10$\,{\rm keV}$
      flux $F_{0.3-10\,{\rm keV}}$ (retrieved for each flux using {\tt cflux}), are shown in Table
      \ref{xfitparams} and the light curve is shown on Figure \ref{FIG:LC}.

\begin{table*}
 \centering
 \begin{minipage}{100mm}
\caption{Results of the spectral fitting of all XRT-PC and PCA observations} 
\begin{tabular}{@{}cccccc@{}}
  \hline
  ObsID & Time & $N_0$ & $\Gamma_X$ & $P(\chi^2)$ & $F_{0.3-10\,{\rm keV}}$\\
               &   MJD-5500   &  ph$\,{\rm cm^{-2}}\,{\rm s^{-1}}$   &                             &         \%               & $\times 10^{-12}\,{\rm
    erg}\,{\rm cm^{-2}}\,{\rm s^{-1}}$ \\\hline
   00036341005 & 247.2-247.2& $(9\pm1) \times10^{-4}$ & $1.62\pm0.14$ & 18 & $6.8\pm0.8$ \\ 
   00036341006 & 249.5-249.7& $(9\pm1) \times10^{-4}$ & $1.45\pm0.09$ & 70 &  $7.6\pm0.6$\\
   00036341007 & 384.7-384.8& $(8.3\pm0.4) \times10^{-4}$ & $1.47\pm0.06$ & 31 & $7.2\pm0.4$\\
   00036341008 & 608.1-608.2& $(10\pm1) \times10^{-4}$ & $1.49^{+0.15}_{-0.14}$ & 35 & $8.3^{+1.0}_{-0.9}$\\
   00036341010 & 608.0-608.0&  & $1.51\pm0.09$ & 94 & $7.8\pm0.6$\\
   00036341011 & 609.8-609.9& $(9.3\pm0.5) \times10^{-4}$ & $1.52^{+0.07}_{-0.06}$ & 60 & $7.6\pm0.4$\\
   sum all above & & $(9.2\pm0.2) \times10^{-4}$ & $1.52\pm0.02$ & $99$ & $7.54\pm0.2$\\
   95141 & 387.9-391.9& $1.3^{+0.4}_{-0.3}\times 10^{-3}$ & $1.74\pm0.16$ & 91 & $5.6\pm0.4$\\
\hline
\end{tabular}
\label{xfitparams} 
\end{minipage}
\end{table*}

Systematic errors on the {\it Swift}-XRT spectra and absolute flux are less than $3\%$ and $10\%$,
respectively \citep{2009A&A...494..775G}, while PCA-XRT cross-calibration details can be found in
\citet{2011A&A...525A..25T}.

\subsection{\chandra\ observations}
\label{sub:chandra}
\begin{figure}
\includegraphics[width=0.49 \textwidth]{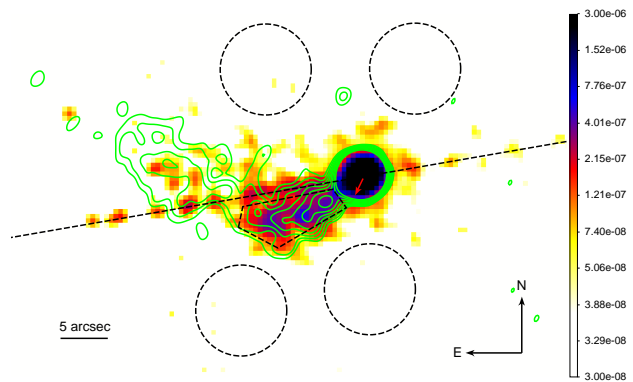}
\caption{Adaptively smoothed, exposure corrected X-ray image obtained by Chandra
in the 0.5--7 keV energy band in units of ph cm$^{-2}$~s$^{-1}$~px$^{-1}$ with
the pixel size of 0.492$\arcsec$, where one arcsecond corresponds to 0.947 kpc
on linear scale. Overlaid are 1.4 GHz radio emission contours at 2$\arcsec$
resolution from reprocessing archival VLA data (program AB700; C.C. Cheung,
private communication). The flux densities of radio contours increase by a
factor of 1.5, starting from a value of 5 times the rms noise equal to $1.82\times
10^{-4}$ Jy beam$^{-1}$. The small red arrow in the radio core shows the orientation of the
milliarcsecond scale radio jet seen in VLBA \citep[see e.g. ][]{1998ApJ...504..702L,2002AJ....124..662Z}, and references
therein, for a discussion on the radio jet at different scales). A black dashed polygon
delimits the region used to calculate the jet spectrum, the dashed circles are the
background regions and a black dashed line indicates a location of an ACIS readout
streak.}
\label{FIG:radiocont}
\end{figure}

AP~Librae was observed by \chandra\ on July 4, 2003 with a total exposure time of 14\,ks. The \chandra\ data
reprocessing and reduction were performed following the standard procedures described in the
\chandra\ Interactive Analysis of Observations\footnote{http://cxc.harvard.edu/ciao/index.html} (CIAO) threads, using CIAO v4.3 and the \chandra\ Calibration
Database (CALDB) version 4.4.6. The data reveal the presence of an extended jet on arcsecond scales, which is
unique amongst the VHE emitting BL Lac class so far. A radio VLA observation was used to align the nuclear X-ray
emission with the radio core. A registered, exposure-corrected and adaptively smoothed image of AP~Librae in
units of ${\rm ph}\,{\rm cm^{-2}}\,{\rm s^{-1}}\,{\rm px^{-1}} $, with radio contours overlaid is shown in
Figure \ref{FIG:radiocont}. In order to assess to what degree the {\it RXTE} and {\it Swift} spectra need
corrections for non-core emission, the spectrum of the jet is estimated, with the caveat that this observation
is not contemporaneous with the data set presented here.

A spectrum of the jet was taken from a polygon shaped region which avoids the emission of the core and the
ACIS readout streak.   A core spectrum comes from a 2 arcsecond region centered on the
core. A background spectrum was extracted from four circular regions placed to the north and
south of the source. The jet and background regions are marked in Figure \ref{FIG:radiocont}. In order to
estimate the effects of pile-up in the core and jet region, the method described by
\cite{2011ApJ...743..177H} was used. In the jet region
no pile-up was found while it was necessary to correct for mild pile-up in the core.

The spectra of the core and the jet contain $\simeq 4900$ and $\simeq 200$ background subtracted counts,
respectively. Both spectra were binned to a minimum of 20 counts per bin, and fit in the 0.5--7.0 keV energy
band using an absorbed power-law model in {\tt XSPEC} with the same $N_{\rm H}$ as in \S\ref{sub:swift}. The fit
of the jet spectrum yields a photon index $\Gamma_{\rm jet}=1.59\pm0.16$ and a 2--10 keV unabsorbed flux of
$F^{\rm jet}_{\rm 2-10 keV}=(1.07\pm0.37)\times10^{-13}$ erg~cm$^{-2}$~s$^{-1}$,  with a $\chi^2=4.4$ for 7 dof, or more than an order of
magnitude below the value measured for the source in \S\ref{sub:swift} based on the {\it Swift} and {\it RXTE} data, which can hence safely be used as the
X-ray flux of the core in AP~Librae. The jet spectrum is comparable with the spectra of large-scale quasar jets observed by \chandra, which may also be sources of relatively intense \gr\ emission \citep[see the discussion in][]{2004ApJ...608..698S,2008ApJ...686..181F}. Such a scenario is not formally excluded here since an extrapolation
of the jet spectrum could connect within the experimental errors with either the HE or VHE fluxes reported
here. Assuming no pile-up, the best power-law fit to the core spectrum yields a
photon index of $\Gamma_{\rm core}=1.51\pm0.03$ and a 2--10 keV unabsorbed flux of $F_{\rm 2-10 keV}^{\rm
  core}=3.18^{+0.19}_{-0.14}\times10^{-12}$ erg~cm$^{-2}$~s$^{-1}$.  Using the {\tt pileup} model in {\tt XSPEC}, a
pile-up corrected spectrum appears however to be softer with $\Gamma_{\rm core}=1.68^{+0.03}_{-0.06}$ and
$F_{\rm 2-10 keV}^{\rm core}\simeq2.31\times10^{-12}$ erg~cm$^{-2}$~s$^{-1}$, with a $\chi^2=158.4$ for 129 dof. The pile-up model of \citet{Davis2001} was used in the fit of the core spectrum, and the value of the pile-up parameter $\alpha > 0$ indicates that the fit is indeed affected by this. However, it was not possible to obtain an
error estimate on $\alpha$, and hence we also do not have an error estimate on the unabsorbed and pile-up
corrected flux. Due to pile-up effects, the fit results for the core should be treated with caution. This extended X-ray jet was first reported by \citet{2013ApJ...776...68K}.  Our results differ slightly, probably because we used different extraction and background regions, and Kaufmann et al. did not take into account the above-mentioned ACIS readout streak.

\subsection{{\it Swift}-UVOT and SMARTS observations}
\label{UV_OPTIC}

All of the available archival data taken on AP~Librae with the
ultra-violet and optical telescope (UVOT) on the {\it{Swift}}
satellite were analyzed. This comprised 35 exposures taken between
April 2007 and July 2010, 13 of which occurred during the time frame
with which this paper is concerned (see Figure \ref{FIG:LC}). After extracting the source counts
from an aperture of 5.0\arcsec radius around AP~Librae and the
background counts from four neighboring regions, each of the same
size, the magnitudes were computed using the
{\it{uvotsource}} tool with calibrations from
  \citet{2011AIPC.1358..373B}. These were converted to fluxes
using the values from \citet{2008MNRAS.383..627P} after correction for
extinction following the procedure and R$_v$ value of
\citet{2009ApJ...690..163R}. The values of $a$ and $b$ from
\citet{2009ApJ...690..163R}, computed following the procedure of
\citet{1989ApJ...345..245C}, were used. The $E(B-V)$ value from
\citet{2011ApJ...737..103S}, accessed through the NASA/IPAC
Extragalactic Database, was used. Results are summarized in Table~\ref{table:SMARTUVOT}.



AP~Librae was observed in context of the Yale {\it Fermi}/SMARTS
project\footnote{http://www.astro.yale.edu/smarts/glast/pubs.html} \citep{Smarts}. Observations
were performed in the B, R, J and K bands between February 27, 2011 (MJD 55619)
and March 3st, 2013 (MJD 56739) and shown on Figure \ref{FIG:LC}. The number of observations and the mean
magnitudes are given in Table~\ref{table:SMARTUVOT} together with the corresponding fluxes.
Magnitudes have been corrected for Galactic absorption using values from
\citet{2011ApJ...737..103S} and converted in flux units using the Bessell zero
points \citep{Bessel90}.

\begin{table*}
 \centering
 \begin{minipage}{100mm}
\caption{Summary of the \textit{Swift}-UVOT and  SMARTS results. Columns 1 and 3 give the filter and corresponding energies and the second column gives the number of observations. Magnitudes (column 4) are not corrected for Galactic absorption. The last column gives the corrected flux.}
\label{table:SMARTUVOT} 
\begin{tabular}{@{}ccccc@{}}
\hline
Filter&$N_{\rm Obs}$&Energy&magnitude&Flux\\
& & [eV] & & [$10^{-11}\ \ergcms$]   

\\\hline 
SMARTS: \\\hline
K & 25 & 0.56 & 11.63$\pm$0.37 & 1.96$\pm$0.78  \\
J & 79 & 0.99 & 12.76$\pm$0.17 & 3.26$\pm$0.54  \\
R & 74 & 1.77 & 14.53$\pm$0.18 & 1.88$\pm$0.34  \\
B & 29 & 2.86 & 15.85$\pm$0.13 & 1.35$\pm$0.18 \\\hline
\textit{Swift}-UVOT: \\\hline
V & 3 & 2.30 & 15.18 $\pm$ 0.04 & 2.43 $\pm$ 0.10 \\
B & 3 & 2.86 & 15.94 $\pm$ 0.04 & 1.69 $\pm$ 0.06 \\
U & 7 & 3.54 & 15.68 $\pm$ 0.04 & 0.96 $\pm$ 0.03 \\
UVW1 & 6 & 4.72 & 15.88 $\pm$ 0.05 & 0.63 $\pm$ 0.03 \\
UVW2 & 10 & 6.12 & 16.12 $\pm$ 0.05 & 0.57 $\pm$ 0.02 \\
UVM2 & 3 & 5.57 &16.09 $\pm$ 0.06 & 0.55 $\pm$ 0.03 \\\hline
\end{tabular}
\end{minipage}
\end{table*}

The host galaxy of AP~Librae is bright and therefore the contribution from
starlight must be taken into account to estimate the non-thermal flux from the
core in the near-infrared to UV band. The dereddened near-infrared and optical
measurements of AP~Librae reported in Figure 1 of \cite{1993AJ....106...11F},
where the total emission was modeled with a giant elliptical galaxy template and
a superposed non-thermal power-law continuum, are given for illustration in the composite SED of Figure
\ref{fig:sedssc}. The synchrotron emission probably peaks in the optical- to
near-IR range, since the spectral index for AP~Librae in that range is
$\alpha_{\rm IROP} = 0.95\pm0.10$. In \cite{2007A&A...476..723H}, the fluxes in
the B and U bands were calculated for the host galaxy and the core. The
fractional contribution of the latter was $\simeq$ 42\% in the B band and
$\simeq$ 69\% in the U band. At higher energies the emission from the core
accounts for an even higher percentage. To take this result into account, the host galaxy template of \citet{1998ApJ...509..103S} has been used and with a normalization ajusted to fit the data.

\subsection{MOJAVE}
\label{mojave}
The parsec-scale structure of the radio jet of AP~Librae has been monitored
throughout the past decade as part of the MOJAVE
program\footnote{http://www.physics.purdue.edu/astro/MOJAVE} (Monitoring of Jets
in Active galactic nuclei with VLBA Experiments) with the Very Long Baseline
Array (VLBA) at a frequency of 15\,GHz. The VLBA data have been calibrated and
analyzed following the procedures described by \citet{2009AJ....137.3718L}. The
source shows a bright, continuous inner jet region with a bright jet core, i.e apparent jet base,
extending towards the South. At a resolution of typically $\simeq (1.5 \times
0.5)$\,milli-arcsecond (mas), the core is not clearly separated from the inner
jet. Elliptical Gaussian components were used to model the brightness
distribution and to determine radio flux densities of different emission regions
within the source. For the comparison with higher-energy multiwavelength data,
we focused on the inner $1.5$\,mas ($\simeq 1.41$\,pc) region, which could
typically be modeled with 2--3 Gaussian model components. We have used different
models with circular and elliptical model components and tested the formal
statistical model-fitting uncertainties of the total flux density, which turn out to be much smaller
($\lesssim (1-3)$\,\%) than the absolute calibration uncertainty, which can be
conservatively estimated to be of the order of $\lesssim 10$\,\%.

The 16 MOJAVE observations from MJD 53853 to 55718 do not show sign of
significante variability in the VLBI core region. Figure \ref{fig:sedssc} shows
the value of 1.48 Jy of the radio flux density, averaged over the full
observations, from the inner 1.5\,mas jet core. 

\section{Discussion}
\label{discussion}

\subsection{The radiative components}
\label{radiativeCompo}

The composite SED of AP~Librae is shown in Figure \ref{fig:sedssc}. Together
with the MOJAVE, SMARTS, \chandra, \textit{Swift}-UVOT, \textit{Swift}-XRT,
\textit{RXTE}, \fla\ and H.E.S.S. data analyzed in this work, archival data from
NED are reported. In the 30--353 GHz band, the \textit{Planck} measurements from
the Early Release Compact Source Catalog \citep[ERCSC,][]{2011A&A...536A...7P} are
in good agreement with the archival data as are the Wide-field Infrared Survey
Explorer \citep[WISE, ][]{WISE} data in the bands 3.4, 4.6, 12, and 22 $\mu$m.

An extrapolation of the hard X-ray to the optical-UV power-law spectrum reported here underestimates the
simultaneous UVOT flux by at least 2 orders of magnitude, though the steeply falling UV spectrum possibly
connects with the onset of the XRT spectrum. This indicates the presence of an inflection point in the SED
widely attributed to a transition from synchrotron to IC dominated radiation. This feature shows that the
Compton component of AP~Librae is the broadest ever observed in {\sl any} blazar, spanning more than 10
decades in energy from $\simeq 0.1\,{\rm keV}$ to $\simeq 1\,{\rm TeV}$. Indeed, only two other objects of the
same class as AP Librae, and hence with broad Compton components, have been detected at VHE energies so far: BL Lacertae ($z=0.069$), the first LBL object to
be proved as being a VHE emitter \citep{2007ApJ...666L..17A}, and S5~0716+714 ($z=0.310$) following an optical
trigger \citep{2009ApJ...704L.129A}. The observed VHE spectrum of the former is not as energetic as AP Librae,
and the X-ray spectrum of the latter appears to still belong to the synchrotron component.

An empirical characterization of the two radiative components, through a third
degree polynomial fit of each hump in $\nu F_\nu$ representation (as in, e.g.,
\citealt{2010ApJ...716...30A}), is used to estimate the synchrotron and IC peak
energies. The values of the parameters obtained from a $\chi^2$ fit\footnote{The EBL absorption has been  taken into account in the fit.} are given in Table \ref{table:PolFit} and the results are
represented in the composite SED of Figure \ref{fig:sedssc}. As mentioned above,
the SMARTS and the \textit{Swift}-UVOT measurements in the V, B and U were not
used in the fit of the synchrotron peak as well as the data from
\citet{1993AJ....106...11F}. The position of the synchrotron peak is then
estimated to be $E_{\rm s,\,peak}\simeq 0.18\pm0.06\,{\rm eV}$, which is
compatible with the value of $E_{\rm s,\,peak}=0.26\,{\rm eV}$ derived by
\citet{2010ApJ...716...30A} on a different data set. The same authors estimated
$E_{\rm ic,\,peak}=2.6^{+3.2}_{-1.4}\,{\rm GeV}$ for AP~Librae in Table 13
based on a strong correlation of $E_{\rm ic,\,peak}$ with the HE photon index
$\Gamma_{\rm HE}$, as expressed in their Equation 5\footnote{The quoted
uncertainty, not given in their table, is derived from their estimation of an
error of 0.7 associated with the estimation of the log of $E_{\rm ic,\,peak}$ in
Equation 5.}. Using the photon index found by \cite{ApLibHESS}, which is
based on an order of magnitude larger data set, yields a lower but still
compatible value of $E_{\rm ic,\,peak}=0.9^{+1.0}_{-0.5}\,{\rm GeV}$ and this value
was constrained to be below 1~GeV by fitting the HE-VHE data \citep{ApLibHESS}.
The polynomial fit presented here yields a much lower value of $E_{\rm
ic,\,peak}=17^{+24}_{-6}\,{\rm MeV}$, which can be attributed to use of the entire SED.
This is the lowest IC component peak ever measured for a TeV-emitting blazar.


\begin{table*}
 \centering
 \begin{minipage}{70mm}
\caption{Parameters of third degree polynomial function describing the low and high-energy component of AP~Librae. The function is of the form $f(E) = p_{0}+p_{1}\log_{10}(E/{\rm eV})+p_{2}\log^2_{10}(E/{\rm eV})+p_{3}\log^3_{10}(E/{\rm eV})$} 
\label{table:PolFit}
  \begin{tabular}{@{}c c c c c c@{}}
  \hline
Energy Range (eV)&$ p_{0}$&$p_{1}$&$p_{2}$&$p_{3}$\\  \hline
$3\times10^{-4}-50$ & -10.79 &-0.52 &-0.41 & -0.048 \\ 
$50-10^{13}$ & -13.36 & 0.82 & -0.068 &-0.001 \\ 
\hline
\end{tabular}
\end{minipage}
\end{table*}

The third degree polynomial also provides a straightforward estimation of the curvatures $\kappa_{\rm s}$ and
$\kappa_{\rm IC}$ at the peak positions $E_{\rm s,\,peak}$ and $E_{\rm ic,\,peak}$, respectively, which
pertain to the widths of each hump. \cite{2009A&A...504..821P} show that a relation $\kappa_{\rm s} =
2\kappa_{\rm IC}$ is expected in a pure Thomson scattering regime, using a logparabolic parametrization of
each of the two humps generated by a single zone homogeneous SSC model, while
$\kappa_{\rm s} = \kappa_{\rm IC}/5$ in the Klein-Nishina (KN) regime. The curvatures found here for AP~Librae
yield a surprising $\kappa_{\rm s} \simeq 6.6 \kappa_{\rm IC}$, emphasizing the broadness of the IC component
compared to the synchrotron hump, which is hardly possible to reproduce with simple radiative models.

\subsection{Radiative scenarios}
\label{SSCnumbers}

In a one zone homogenous SSC framework, electrons produce synchrotron photons which
are upscattered through the IC mechanism by the same electrons to generate the HE and VHE
photons. If this upscattering occurs predominantly in the Thomson regime up to the peak energy,
then it becomes possible to constrain the product of the magnetic field $B$ and the Doppler factor $\delta$
for a single zone homogenous SSC model (following \citealt{1998ApJ...509..608T}, Equation~4):
\begin{equation}
B\delta = (1+z) \frac{8.6\times10^7{E}^{2}_{\rm s,\,peak }}{E_{\rm ic,\,peak}},
\label{eq2}
\end{equation}
where the peak energies are expressed in eV. Using the range for $E_{\rm s,\,peak}$ found previously and $E_{\rm ic,\,peak}=$~17~MeV yields $B \delta = 0.17\,{\rm G}$. The value of the
break Lorentz factor $\gamma_b$ of the underlying electron distribution can also be derived from the ratio of the peak emission energies as
 $\sqrt{\frac{3E_{\rm ic,\,peak}}{4E_{\rm s,\,peak}}}\simeq 8.5\times10^3 $.


Assuming now that the observed synchrotron radiation does not exceed $\simeq 0.1\,{\rm
  keV}$, (i.e. the lowest energy bin in the XRT
spectrum), which is more likely to belong to the onset of the IC component, then this constrains the maximal
Lorentz factor $\gamma_{\rm max}$ of the underlying electron population through the maximum synchrotron energy
$$E_{\rm s,\, max} \simeq \gamma_{\rm max}^2\frac{B\delta \mec}{B_{\rm cr}(1+z)} \leq 0.1\,{\rm keV},$$
where $B_{\rm cr}=4.414\times10^{13}$G is the critical magnetic field leading to
\begin{equation}
\gamma_{\rm max} \leq 10^5 B^{-1/2} \delta^{-1/2}.
\label{eq1}
\end{equation}
Using Equation\,\ref{eq1} and Equation.\,\ref{eq2} then yields $\gamma_{\rm max}\leq 2.4\times10^5$, which is consistent with
being a factor $\sqrt{E_{\rm s,\, max}/E_{\rm s,\,peak}}$ higher than $\gamma_b$ as expected.


Supposing that electrons with an apparent energy of $\delta\gamma_{\rm max}$ have sufficient energy to upscatter
photons to at least the maximal observed Compton energy $E_{\rm ic,\,max}\simeq 1\,{\rm TeV}$, then the Doppler factor is constrained to a
reasonable value of $\delta \geq 10$.  If the scattering of 0.1 keV photons occurs in the Thomson regime, the
Doppler factor should be such that $4\gamma_{\rm max}\times 0.1\,{\rm keV}\leq\delta\mec$. Using the value for
$\gamma_{\rm max}$ found above leads to an unusually high value of $\delta\geq163$.  If however the scattering occurs
in the KN regime for these highest energy seed photons, then $E_{\rm ic,\,max}= \frac{\delta\mec}{1+z}
\gamma_{\rm max}$ which, combined with the above constraint (Equation\,\ref{eq1}) on $\gamma_{\rm max}$, then yields
\begin{equation}
B\delta^{-1} \leq 2.3\times 10^{-3}\,{\rm G},
\label{eq4}
\end{equation}
from which follows, using the above constraint $\delta\geq 10$, a
reasonable constraint of $B \leq 2.3\times10^{-2}\,{\rm G}$. In Appendix
\ref{appendix}, similar conclusions are drawn for an arbitrary type of seed photons. 

\citet{2013AJ....146..120L} measured an apparent superluminal motion 6.4c. This is compatible with $\delta > 10$ for a viewing angle below $< 5$ degrees and with $\delta = 20$ for 1.7 degrees.


Going further by assuming that photons with energies up to $E_{\rm ic,\,peak}$
are produced in the Thomson regime, and the $\simeq 1\,{\rm TeV}$ photons in the
KN regime, then Equation\,\ref{eq2} and Equation\,\ref{eq4} can be combined to give $B \leq
2\times 10^{-2}\,{\rm G}$ regardless of the value of $\delta$.


\subsection{Application of an SSC model to the SED}

The time-averaged SED of AP~Librae was modeled with a canonical one zone
homogeneous SSC model \citep{THEO::SSC_BAND}. A spherical region of size $R$,
with an electron distribution $N_e(\gamma)$, moving with a bulk Doppler factor
$\delta$, is filled uniformly with a magnetic field $B$. As in
\citet{2010MNRAS.401.1570T}, $N_e(\gamma)$ is described by a broken power-law of
index $S_1$ between $\gamma=1$ and $\gamma_{\rm b}$ and $S_2$ between
$\gamma_{\rm b}$ and $\gamma_{\rm max}$. The electrons lose their energy by
synchrotron emission, producing a field of photons which become the targets for
the same electron population through the IC process. The KN effects are taken
into account using the \textit{Jones kernel} \citep{JONES_KERNEL} to compute the
IC cross-section. 

\begin{figure}
\includegraphics[width=0.49 \textwidth]{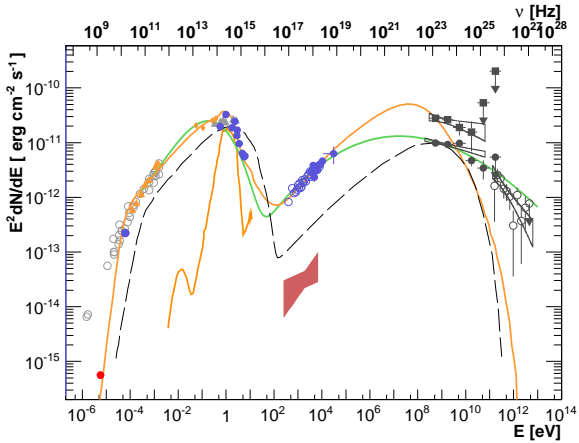}
\caption{The broadband SED of the LBL AP~Librae. The orange triangles come from the \textit{Planck} ERCSC. The orange diamonds are WISE measurements. Blue points are, from low energy to high energy, MOJAVE(15GHz) , \textit{Swift}-UVOT (2.30--5.57 eV), SMARTS (0.56--2.86 eV), \textit{Swift}-XRT/\textit{RXTE} (3--50 keV). Grey points and butterflies are \fla\ for the quiet (circle)  and flare (square)  periods (0.3--300 GeV) and H.E.S.S. (E$>$100GeV) measurement from \citep{ApLibHESS}. \textit{Swift}-UVOT, SMARTS data are corrected for Galactic extinction and X-ray data are corrected for $N_{\rm H}$ absorption. Light gray data are taken from NED. The dark gray triangles come from \citet{1993AJ....106...11F}. The red point is the radio flux of the extended jet. The orange line is the host galaxy template of \citet{1998ApJ...509..103S}. The fit with two third degree polynomial functions, not corrected for EBL, are shown with a green line (see \S\ref{radiativeCompo}). The red butterfly is the \chandra\, spectrum from the jet. The dashed line is the SSC model from \citet{2010MNRAS.401.1570T} whereas the red line is the model obtained in this work (see Table \ref{SSCparams}).}
\label{fig:sedssc}
\end{figure}

A tentative model is shown in Figure \ref{fig:sedssc}, where the shape of the
electron distribution ($S_1$, $S_2$ and $\gamma_{\rm b}$) is constrained by the
observed synchrotron component. The remaining parameters ($R$, $B$, $\delta$,
and the total number of electrons $N_{\rm e,\, tot} $) are adjusted to reproduce the
onset of the Compton component in the X-rays. The obtained parameters and model curves, as given
in Table \ref{SSCparams} and Figure \ref{fig:sedssc}, respectively (together with the model parameters and curves derived by \citet{2010MNRAS.401.1570T} for comparison) obey the constraints found in
\S\ref{SSCnumbers}. Not surprisingly, the broad IC component of the SED is difficult to
reconcile with the synchrotron distribution using such a simple model, for which
strong indications were already presented in \S\ref{radiativeCompo}.

The SSC calculation reproduces well the lower energy part of the SED, up to the
X-rays, but the spectral prediction in the \fla\, energy range is much softer,
as well as about one order of magnitude above the observed HE flux. The direct
consequence of the broadness of the IC component is that the \hess\ flux is
largely underestimated. Directly linked to the electron distribution and to the
well measured synchrotron component, this shape can only be affected by the KN
effects, which tend to soften the spectrum leading inevitably to even larger
disagreements. 



\begin{table*}
 \centering
 \begin{minipage}{80mm}
\caption{Parameters of the SSC model presented in this work and from \citet{2010MNRAS.401.1570T}. For both models, $\gamma_{\rm min} = 1 $ was used.} 
\label{SSCparams}
  \begin{tabular}{@{}ccccccccc@{}}
  \hline
Model &$ \gamma_{\rm b}$&$\gamma_{\rm max}$ &S$_1$&S$_2$&$N_{\rm e,\,tot}$&$B$&$R$&$\delta$\\  \hline
&$10^4$&$10^4$ && &$10^{53} $&$10^{-2}$[G]&$10^{16}$[cm] &\\
This work       & 1.1& $ 2.3 $ & 2 & 4.9 &  5.4 & 0.9   & 3.5 & 20  \\ 
Tavecchio et al & 2.0 & 5 & 2 & 4.9 & 0.4 & 1.2 & 1 & 40 \\ 
\hline
\end{tabular}
\end{minipage}
\end{table*}

\subsection{VHE $\gamma$ Rays from the extended jet?}

\begin{figure}
\includegraphics[width=0.49 \textwidth]{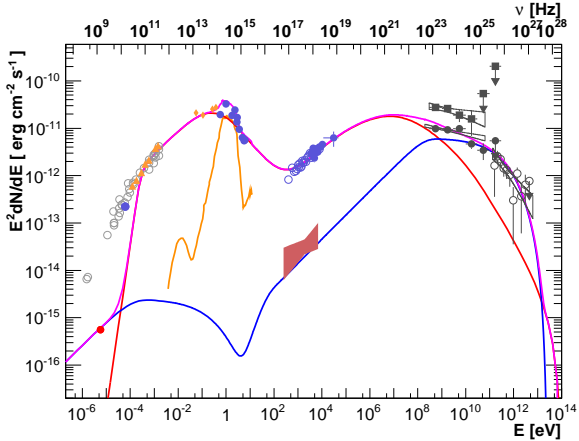}
\caption{Same as Fig.~\ref{fig:sedssc}. The red line is the results of the SSC model from the compact component and the blue line is the flux originating from the extended jet; parameters are given in Table \ref{table_fit}. Purple line is the sum of both.}
\label{sedjustin}
\end{figure}

As seen in the previous sections, one-zone SSC models cannot reproduce the
broadband SED of AP~Librae.  However, \citet*{boett08} proposed
that the Compton-scattering of the cosmic microwave background (CMB)
by electrons in an extended kpc-scale jet could make VHE
$\gamma$-rays.  This model was suggested to explain the hard VHE
spectrum from 1ES~1101$-$232 as observed by \hess\ 
\citep{2006Natur.440.1018A,2007A&A...470..475A}, when EBL attenuation was taken into
account with the models available at the time.  AP~Librae has an extended
kpc-scale jet resolved in radio (see Figures \ref{fig:sedssc} and \ref{sedjustin}) and X-rays (see section \ref{sub:chandra}), and it
has long been thought that the Compton-scattering of CMB photons could
produce the X-rays observed from these extended jets
\citep[e.g.,][]{2000ApJ...544L..23T,2001MNRAS.321L...1C}.  Therefore, it seems natural
to apply this model to AP~Librae, to see if the extended jet emission
could plausibly make up the VHE $\gamma$ rays.  Thus, the
broadband SED of AP~Librae has been modeled with a compact, synchrotron/SSC model based on
\citet{2008ApJ...686..181F}, and an additional component from the extended
jet, emitting synchrotron and inverse Compton-scattering of CMB
photons (hereafter ICCMB).

The result of this model is shown in Figure \ref{sedjustin}, with the
model parameters in Table \ref{table_fit}.  The model parameters are
fully described in \citet{2008ApJ...686..181F}.  The compact component can
explain the radio, optical (not including emission that is clearly
from the host galaxy), X-ray, and the lower-energy {\it Fermi}-LAT
$\gamma$-ray data.  The extended component can explain the extended
radio and X-ray data, as well as the highest $\g$-ray emission
detected by the LAT and H.E.S.S.  A double-broken power-law was used to
describe the electron distribution in the compact component, while
only a single broken power-law was needed for the electron
distribution in the extended component.  Parameters in the compact
component are broadly comparable to synchrotron/SSC modeling results
for other BL Lac objects, including the jet power in electrons being
several orders of magnitude greater than that in the magnetic field
\citep[e.g.,][]{2008ApJ...686..181F, 2011ApJ...736..131A,
2011ApJ...727..129A,2011ApJ...730..101A, 2011ApJ...726...43A, 2013ApJ...779...92A, 2014ApJ...782...13A,
2014ApJ...797...89A}.  The extended jet is much closer to equipartition
between electron and magnetic field density by design; a model out of
equipartition would still be able to reproduce the data.  These parameters
are also close to previous results for modeling extended jets,
although the magnetic field is a bit lower than usual
\citep[typically found $>1\ \mu\Gauss$; e.g.,][]{2007ApJ...662..900T}.  This
may be because previous ICCMB models of extended jets are for FSRQs, rather than
BL Lac objects. One hypothesis can be that the magnetic fields in extended jets
of BL Lac objects are lower than those in the extended jets of FSRQs.

It should be noted that the ICCMB model for explaining the X-ray
emission from extended jets is controversial.  It could be that X-rays
are instead produced by synchrotron emission from another population of
electrons in the extended jet \citep[e.g.,][]{2004ApJ...613..151A,2006MNRAS.366.1465H}. In this alternative framework, HE and VHE emission is unlikely.
Recently, \citet{2014ApJ...780L..27M} used {\it Fermi}-LAT observations to rule
out the ICCMB model for the X-ray emission from the extended jet in
the FSRQ 3C 273.

\begin{table*}
 \centering
 \begin{minipage}{100mm}
\caption{Model parameters for the SED shown in Fig.~\ref{sedjustin}. The redshift $z$ is 0.049.} 
\label{table_fit} 
  \begin{tabular}{@{}lccc@{}}
  \hline
Parameter&Symbol&Compact component&Extended Jet\\

Bulk Lorentz Factor & $\Gamma$	& 20	& 8 \\
Doppler factor & $\delta_D$	& 20	& 8 \\
Magnetic Field [G]& $B$         & 0.05  & $5.6\times10^{-7}$ \\
Variability Timescale [s]& $t_v$ & $3.0\times10^4$ & $1.35\times10^{11}$ \\
Comoving radius of blob [cm]& $R^{\prime}_b$ & $1.7\times10^{16}$ & $3.08\times10^{22}$      \\
\hline
First Electron Spectral Index   & $p_1$       & 2.0 & 2.0 \\
Second Electron Spectral Index  & $p_2$       & 3.0 & 4.0 \\
Third Electron Spectral Index   & $p_3$       & 4.2 &     \\
Minimum Electron Lorentz Factor & $\gamma^{\prime}_{min}$  & $1.0$ & $2.0$ \\
Break Electron Lorentz Factor 1 & $\gamma^{\prime}_{brk,1}$ & $2.8\times10^3$ & $4.9\times10^4$ \\
Break Electron Lorentz Factor 2 & $\gamma^{\prime}_{brk,2}$ & $6.8\times10^3$ &                 \\
Maximum Electron Lorentz Factor & $\gamma^{\prime}_{max}$  & $1.0\times10^7$ & $2.0\times10^6$  \\
\hline
Jet Power in Magnetic Field [$\erg\ \s^{-1}$] & $P_{j,B}$ & $2.2\times10^{42}$ & $1.4\times10^{44}$ \\
Jet Power in Electrons [$\erg\ \s^{-1}$] & $P_{j,e}$ & $1.7\times10^{45}$ & $2.8\times10^{44}$ \\

\hline
\end{tabular}
\end{minipage}
\end{table*}

\subsection{Comparison with other LBL objects}
\label{subsect:comp}

The SEDs of LBL objects detected in VHE $\gamma$-rays challenge single zone
homogeneous SSC radiative models, which usually reproduce reasonably well the
time-averaged SEDs of the HBL class.

The most complete simultaneous coverage of the BL Lacertae was established by \citet{2011ApJ...730..101A}
during a multi-wavelength campaign including the \fla\, and the X-ray observatories mentioned in this study
for the high-energy part. The X-ray spectrum during that campaign was soft, indicating that its origin
was synchrotron radiation rather than Comptonized photons, making for a wider synchrotron $\nu F_\nu$
distribution than is reported here for AP~Librae. The difficulty in this case for modeling BL Lacertae was
that the simulated SED required the energy densities to be far from equipartition. However, a 1997 {\it
  Beppo}-SAX observation \citep{2002A&A...383..763R} of BL Lacertae showed a clear IC origin for the X-ray
radiation, yielding a narrower synchrotron distribution, for which the SSC model failed to reproduce a
reasonable (non-simultaneous) HE spectrum, and an external contribution was added. 

The broad Compton distribution of S5~0716+714, with emission up to
$\simeq 700\,{\rm GeV}$, is either an order of magnitude {\sl below} the best SSC
model prediction from \citet{2009ApJ...704L.129A}, or is too wide if the \fla\,
spectrum constrains the flux at $E_{\rm ic,\,peak}$ (Figure 6 in
\citealt{2010MNRAS.401.1570T}; see also the similar situation for BL Lacertae in
the same Figure). Note that the HE and VHE data were not taken simultaneously in
these two LBL objects.

\section{Conclusions}

Contemporaneous observations of AP~Librae with many currently available space-
and ground-based instruments have been presented. The data have revealed the broadest Compton
distribution of any known blazar to date, which spans from X-ray to TeV energies. 

The SED of AP~Librae is difficult to reproduce with a single zone SSC model: the
steep UV spectrum, probably synchrotron emission, does not connect smoothly with
the X-ray spectrum, which is underestimated by an order of magnitude if a match
is required with the HE $\gamma$-ray spectrum (as was also pointed out by
\citealt{2010MNRAS.401.1570T}). If a match is required with the X-rays, the
\fla\, spectrum is then largely overestimated. The new \hess\ spectrum further
complicates the situation, as none of the previous constraints allows this SSC
model to reach the VHE domain, even assuming a predominantly Thomson scattering
regime which yields Compton components roughly twice as large in $\nu F_\nu$ as
the synchrotron component. There are ways out of the conundrum but at the cost of
increased model complexity. An example is blob-in-jet model, recently proposed by
\citet{2015arXiv150301377H} to reproduce the SED of Ap Librae. Another
possiblility is a model where electrons also upscatter soft photons originating
outside of the high-energy emission site. It has been shown in this work that
VHE $\gamma$ rays from the extended jet, seen in X-ray, can be produced and can
explain the \hess\ spectrum.


AP~Librae is the third of VHE
detected LBL-type object for which single zone SSC models fail to reproduce the
SED, and is currently the only BL Lac type object combining VHE emission and a
resolved X-ray jet.  The LBL class of VHE emitting objects proves to be an
interesting laboratory to test radiative model scenarios, and perhaps to
identify parameters on which the LBL-HBL sequence could depend.

\section*{Acknowledgements}

The \textit{Fermi} LAT Collaboration acknowledges generous ongoing support
from a number of agencies and institutes that have supported both the
development and the operation of the LAT as well as scientific data analysis.
These include the National Aeronautics and Space Administration and the
Department of Energy in the United States, the Commissariat \`a l'Energie Atomique
and the Centre National de la Recherche Scientifique / Institut National de Physique
Nucl\'eaire et de Physique des Particules in France, the Agenzia Spaziale Italiana
and the Istituto Nazionale di Fisica Nucleare in Italy, the Ministry of Education,
Culture, Sports, Science and Technology (MEXT), High Energy Accelerator Research
Organization (KEK) and Japan Aerospace Exploration Agency (JAXA) in Japan, and
the K.~A.~Wallenberg Foundation, the Swedish Research Council and the
Swedish National Space Board in Sweden.
 
Additional support for science analysis during the operations phase is gratefully acknowledged from the Istituto Nazionale di Astrofisica in Italy and the Centre National d'\'Etudes Spatiales in France.

This research has made use of data from the MOJAVE database that is maintained
by the MOJAVE team (Lister et al. 2009). The MOJAVE program is supported under NASA-Fermi grant NNX12A087G.

The National Radio Astronomy Observatory is a facility of the National
Science Foundation operated under cooperative agreement by Associated
Universities, Inc.
 
This research has made use of the NASA/IPAC Extragalactic Database (NED) 
which is operated by the Jet Propulsion Laboratory, California Institute of Technology, 
under contract with the National Aeronautics and Space Administration.

This publication makes use of data products from the Wide-field Infrared Survey Explorer, which is a joint project of the University of California, Los Angeles, and the Jet Propulsion Laboratory/California Institute of Technology, funded by the National Aeronautics and Space Administration.

The authors want to acknowledge C.C.~Cheung for the VLA radio
observation used for contours presented in Figure
\ref{FIG:radiocont}. A.S. acknowledges useful discussions with Dan Harris on the problematics of \chandra\ data analyses. We thank the {\rm Swift} and {\em RXTE} teams for their
cooperation in joint observations of AP~Librae. This research has made use of
data provided by the SIMBAD database, operated at CDS, Strasbourg, France. 

D.S. was partially supported by the Labex ENIGMASS

YYK and ABP were supported in part by the Russian Foundation for Basic Research (project 13-02-12103) 

TS was partly supported by the Academy of Finland project 274477.

\bibliography{aplib_MWL_v1}

\begin{thebibliography}{66}
\expandafter\ifx\csname natexlab\endcsname\relax\def\natexlab#1{#1}\fi

\bibitem[{{Abdo} {et~al.}(2010){Abdo}, {Ackermann}, {Agudo}, {Ajello}, {Aller},
  {Aller}, {Angelakis}, {Arkharov}, {Axelsson}, {Bach}, \&
  et~al.}]{2010ApJ...716...30A}
{Abdo}, A.~A., {et~al.} 2010, \apj, 716, 30

\bibitem[{{Abdo} {et~al.}(2011{\natexlab{a}}){Abdo}, {Ackermann}, {Ajello},
  {Baldini}, {Ballet}, {Barbiellini}, {Bastieri}, {Bechtol}, {Bellazzini},
  {Berenji}, \& et~al.}]{2011ApJ...736..131A}
---. 2011{\natexlab{a}}, \apj, 736, 131

\bibitem[{{Abdo} {et~al.}(2011{\natexlab{b}}){Abdo}, {Ackermann}, {Ajello},
  {Allafort}, {Baldini}, {Ballet}, {Barbiellini}, {Baring}, {Bastieri},
  {Bechtol}, \& et~al.}]{2011ApJ...727..129A}
---. 2011{\natexlab{b}}, \apj, 727, 129

\bibitem[{{Abdo} {et~al.}(2011{\natexlab{c}}){Abdo}, {Ackermann}, {Ajello},
  {Baldini}, {Ballet}, {Barbiellini}, {Bastieri}, {Bechtol}, {Bellazzini},
  {Berenji}, \& et~al.}]{2011ApJ...726...43A}
---. 2011{\natexlab{c}}, \apj, 726, 43

\bibitem[{{Abdo} {et~al.}(2011{\natexlab{d}}){Abdo}, {Ackermann}, {Ajello},
  {Antolini}, {Baldini}, {Ballet}, {Barbiellini}, {Bastieri}, {Bechtol},
  {Bellazzini}, {Berenji}, {Blandford}, {Bonamente}, {Borgland}, {Bregeon},
  {Brez}, {Brigida}, {Bruel}, {Buehler}, {Buson}, {Caliandro}, {Cameron},
  {Cannon}, {Caraveo}, {Carrigan}, {Casandjian}, {Cecchi}, {{\c C}elik},
  {Charles}, {Chekhtman}, {Cheung}, {Chiang}, {Ciprini}, {Claus},
  {Cohen-Tanugi}, {Conrad}, {Costamante}, {Cutini}, {Dermer}, {de Palma},
  {Donato}, {Silva}, {Drell}, {Dubois}, {Escande}, {Favuzzi}, {Fegan}, {Finke},
  {Focke}, {Fortin}, {Frailis}, {Fukazawa}, {Funk}, {Fusco}, {Gargano},
  {Gasparrini}, {Gehrels}, {Germani}, {Giglietto}, {Giordano}, {Giroletti},
  {Glanzman}, {Godfrey}, {Grenier}, {Guiriec}, {Hadasch}, {Hayashida}, {Hays},
  {Hughes}, {Itoh}, {J{\'o}hannesson}, {Johnson}, {Johnson}, {Kamae},
  {Katagiri}, {Kataoka}, {Kn{\"o}dlseder}, {Kuss}, {Lande}, {Larsson},
  {Latronico}, {Lee}, {Llena Garde}, {Longo}, {Loparco}, {Lott}, {Lovellette},
  {Lubrano}, {Makeev}, {Mazziotta}, {McEnery}, {Mehault}, {Michelson},
  {Mizuno}, {Monte}, {Monzani}, {Morselli}, {Moskalenko}, {Murgia}, {Nakamori},
  {Naumann-Godo}, {Nishino}, {Nolan}, {Norris}, {Nuss}, {Ohsugi}, {Okumura},
  {Omodei}, {Orlando}, {Ormes}, {Ozaki}, {Paneque}, {Panetta}, {Parent},
  {Pelassa}, {Pepe}, {Pesce-Rollins}, {Piron}, {Porter}, {Rain{\`o}}, {Rando},
  {Razzano}, {Reimer}, {Reimer}, {Ritz}, {Roth}, {Sadrozinski}, {Sanchez},
  {Sander}, {Schinzel}, {Sgr{\`o}}, {Siskind}, {Smith}, {Sokolovsky},
  {Spandre}, {Spinelli}, {Strickman}, {Suson}, {Takahashi}, {Tanaka}, {Thayer},
  {Thayer}, {Thompson}, {Tibaldo}, {Torres}, {Tosti}, {Tramacere}, {Uehara},
  {Usher}, {Vandenbroucke}, {Vasileiou}, {Vilchez}, {Vitale}, {Waite},
  {Wallace}, {Wang}, {Winer}, {Wood}, {Yang}, {Ylinen}, {Ziegler}, {Berdyugin},
  {Boettcher}, {Carrami{\~n}ana}, {Carrasco}, {de la Fuente}, {Diltz},
  {Hovatta}, {Kadenius}, {Kovalev}, {L{\"a}hteenm{\"a}ki}, {Lindfors},
  {Marscher}, {Nilsson}, {Pereira}, {Reinthal}, {Roustazadeh}, {Savolainen},
  {Sillanp{\"a}{\"a}}, {Takalo}, \& {Tornikoski}}]{2011ApJ...730..101A}
---. 2011{\natexlab{d}}, \apj, 730, 101

\bibitem[{{Abramowski} {et~al.}(2015){Abramowski}, {Acero}, {Aharonian},
  {Akhperjanian}, {Anton}, {Balenderan}, {Balzer}, {Barnacka}, {Becherini},
  {Becker Tjus}, {Bernl{\"o}hr}, {Birsin}, {Biteau}, {Bochow}, {Boisson},
  {Bolmont}, {Bordas}, {Brucker}, {Brun}, {Brun}, {Bulik}, {Carrigan},
  {Casanova}, {Cerruti}, {Chadwick}, {Charbonnier}, {Chaves}, {Cheesebrough},
  {Cologna}, {Conrad}, {Couturier}, {Dalton}, {Daniel}, {Davids}, {Degrange},
  {Deil}, {deWilt}, {Dickinson}, {Djannati-Ata{\"i}}, {Domainko}, {O'C.~Drury},
  {Dubus}, {Dutson}, {Dyks}, {Dyrda}, {Egberts}, {Eger}, {Espigat}, {Fallon},
  {Farnier}, {Fegan}, {Feinstein}, {Fernandes}, {Fernandez}, {Fiasson},
  {Fontaine}, {F{\"o}rster}, {F{\"u}{\ss}ling}, {Gajdus}, {Gallant},
  {Garrigoux}, {Gast}, {Giebels}, {Glicenstein}, {Gl{\"u}ck}, {G{\"o}ring},
  {Grondin}, {H{\"a}ffner}, {Hague}, {Hahn}, {Hampf}, {Harris}, {Heinz},
  {Heinzelmann}, {Henri}, {Hermann}, {Hillert}, {Hinton}, {Hofmann},
  {Hofverberg}, {Holler}, {Horns}, {Jacholkowska}, {Jahn}, {Jamrozy}, {Jung},
  {Kastendieck}, {Katarzy{\'n}ski}, {Katz}, {Kaufmann}, {Kh{\'e}lifi},
  {Klochkov}, {Klu{\'z}niak}, {Kneiske}, {Komin}, {Kosack}, {Kossakowski},
  {Krayzel}, {Laffon}, {Lamanna}, {Lenain}, {Lennarz}, {Lohse}, {Lopatin},
  {Lu}, {Marandon}, {Marcowith}, {Masbou}, {Maurin}, {Maxted}, {Mayer},
  {McComb}, {Medina}, {M{\'e}hault}, {Menzler}, {Moderski}, {Mohamed},
  {Moulin}, {Naumann}, {Naumann-Godo}, {de Naurois}, {Nedbal}, {Nguyen},
  {Niemiec}, {Nolan}, {Ohm}, {de O{\~n}a Wilhelmi}, {Opitz}, {Ostrowski},
  {Oya}, {Panter}, {Parsons}, {Paz Arribas}, {Pekeur}, {Pelletier}, {Perez},
  {Petrucci}, {Peyaud}, {Pita}, {P{\"u}hlhofer}, {Punch}, {Quirrenbach},
  {Raue}, {Reimer}, {Reimer}, {Renaud}, {de los Reyes}, {Rieger}, {Ripken},
  {Rob}, {Rosier-Lees}, {Rowell}, {Rudak}, {Rulten}, {Sahakian}, {Sanchez},
  {Santangelo}, {Schlickeiser}, {Schulz}, {Schwanke}, {Schwarzburg},
  {Schwemmer}, {Sheidaei}, {Skilton}, {Sol}, {Spengler}, {Stawarz},
  {Steenkamp}, {Stegmann}, {Stinzing}, {Stycz}, {Sushch}, {Szostek},
  {Tavernet}, {Terrier}, {Tluczykont}, {Valerius}, {van Eldik}, {Vasileiadis},
  {Venter}, {Viana}, {Vincent}, {V{\"o}lk}, {Volpe}, {Vorobiov}, {Vorster},
  {Wagner}, {Ward}, {White}, {Wierzcholska}, {Wouters}, {Zacharias}, {Zajczyk},
  {Zdziarski}, {Zech}, \& {Zechlin}}]{ApLibHESS}
{Abramowski}, A., {et~al.} 2015, \aap, 573, A31

\bibitem[{{Ackermann} {et~al.}(2011){Ackermann}, {Ajello}, {Allafort},
  {Antolini}, {Atwood}, {Axelsson}, {Baldini}, {Ballet}, {Barbiellini},
  {Bastieri}, {Bechtol}, {Bellazzini}, {Berenji}, {Blandford}, {Bloom},
  {Bonamente}, {Borgland}, {Bottacini}, {Bouvier}, {Bregeon}, {Brigida},
  {Bruel}, {Buehler}, {Burnett}, {Buson}, {Caliandro}, {Cameron}, {Caraveo},
  {Casandjian}, {Cavazzuti}, {Cecchi}, {Charles}, {Cheung}, {Chiang},
  {Ciprini}, {Claus}, {Cohen-Tanugi}, {Conrad}, {Costamante}, {Cutini}, {de
  Angelis}, {de Palma}, {Dermer}, {Digel}, {Silva}, {Drell}, {Dubois},
  {Escande}, {Favuzzi}, {Fegan}, {Ferrara}, {Finke}, {Focke}, {Fortin},
  {Frailis}, {Fukazawa}, {Funk}, {Fusco}, {Gargano}, {Gasparrini}, {Gehrels},
  {Germani}, {Giebels}, {Giglietto}, {Giommi}, {Giordano}, {Giroletti},
  {Glanzman}, {Godfrey}, {Grenier}, {Grove}, {Guiriec}, {Gustafsson},
  {Hadasch}, {Hayashida}, {Hays}, {Healey}, {Horan}, {Hou}, {Hughes},
  {Iafrate}, {J{\'o}hannesson}, {Johnson}, {Johnson}, {Kamae}, {Katagiri},
  {Kataoka}, {Kn{\"o}dlseder}, {Kuss}, {Lande}, {Larsson}, {Latronico},
  {Longo}, {Loparco}, {Lott}, {Lovellette}, {Lubrano}, {Madejski}, {Mazziotta},
  {McConville}, {McEnery}, {Michelson}, {Mitthumsiri}, {Mizuno}, {Moiseev},
  {Monte}, {Monzani}, {Moretti}, {Morselli}, {Moskalenko}, {Murgia},
  {Nakamori}, {Naumann-Godo}, {Nolan}, {Norris}, {Nuss}, {Ohno}, {Ohsugi},
  {Okumura}, {Omodei}, {Orienti}, {Orlando}, {Ormes}, {Ozaki}, {Paneque},
  {Parent}, {Pesce-Rollins}, {Pierbattista}, {Piranomonte}, {Piron}, {Pivato},
  {Porter}, {Rain{\`o}}, {Rando}, {Razzano}, {Razzaque}, {Reimer}, {Reimer},
  {Ritz}, {Rochester}, {Romani}, {Roth}, {Sanchez}, {Sbarra}, {Scargle},
  {Schalk}, {Sgr{\`o}}, {Shaw}, {Siskind}, {Spandre}, {Spinelli}, {Strong},
  {Suson}, {Tajima}, {Takahashi}, {Takahashi}, {Tanaka}, {Thayer}, {Thayer},
  {Thompson}, {Tibaldo}, {Tinivella}, {Torres}, {Tosti}, {Troja}, {Uchiyama},
  {Vandenbroucke}, {Vasileiou}, {Vianello}, {Vitale}, {Waite}, {Wallace},
  {Wang}, {Winer}, {Wood}, {Wood}, \& {Zimmer}}]{2011ApJ...743..171A}
{Ackermann}, M., {et~al.} 2011, \apj, 743, 171

\bibitem[{{Aharonian} {et~al.}(2006){Aharonian}, {Akhperjanian}, {Bazer-Bachi},
  {Beilicke}, {Benbow}, {Berge}, {Bernl{\"o}hr}, {Boisson}, {Bolz}, {Borrel},
  {Braun}, {Breitling}, {Brown}, {Chadwick}, {Chounet}, {Cornils},
  {Costamante}, {Degrange}, {Dickinson}, {Djannati-Ata{\"\i}}, {Drury},
  {Dubus}, {Emmanoulopoulos}, {Espigat}, {Feinstein}, {Fontaine}, {Fuchs},
  {Funk}, {Gallant}, {Giebels}, {Gillessen}, {Glicenstein}, {Goret},
  {Hadjichristidis}, {Hauser}, {Hauser}, {Heinzelmann}, {Henri}, {Hermann},
  {Hinton}, {Hofmann}, {Holleran}, {Horns}, {Jacholkowska}, {de Jager},
  {Kh{\'e}lifi}, {Klages}, {Komin}, {Konopelko}, {Latham}, {Le Gallou},
  {Lemi{\`e}re}, {Lemoine-Goumard}, {Leroy}, {Lohse}, {Martin},
  {Martineau-Huynh}, {Marcowith}, {Masterson}, {McComb}, {de Naurois}, {Nolan},
  {Noutsos}, {Orford}, {Osborne}, {Ouchrif}, {Panter}, {Pelletier}, {Pita},
  {P{\"u}hlhofer}, {Punch}, {Raubenheimer}, {Raue}, {Raux}, {Rayner}, {Reimer},
  {Reimer}, {Ripken}, {Rob}, {Rolland}, {Rowell}, {Sahakian}, {Saug{\'e}},
  {Schlenker}, {Schlickeiser}, {Schuster}, {Schwanke}, {Siewert}, {Sol},
  {Spangler}, {Steenkamp}, {Stegmann}, {Tavernet}, {Terrier}, {Th{\'e}oret},
  {Tluczykont}, {van Eldik}, {Vasileiadis}, {Venter}, {Vincent}, {V{\"o}lk}, \&
  {Wagner}}]{2006Natur.440.1018A}
{Aharonian}, F., {et~al.} 2006, \nat, 440, 1018

\bibitem[{{Aharonian} {et~al.}(2007){Aharonian}, {Akhperjanian}, {Bazer-Bachi},
  {Beilicke}, {Benbow}, {Berge}, {Bernl{\"o}hr}, {Boisson}, {Bolz}, {Borrel},
  {Braun}, {Brion}, {Brown}, {B{\"u}hler}, {B{\"u}sching}, {Boutelier},
  {Carrigan}, {Chadwick}, {Chounet}, {Coignet}, {Cornils}, {Costamante},
  {Degrange}, {Dickinson}, {Djannati-Ata{\"i}}, {O'C.~Drury}, {Dubus},
  {Egberts}, {Emmanoulopoulos}, {Espigat}, {Farnier}, {Feinstein}, {Ferrero},
  {Fiasson}, {Fontaine}, {Funk}, {Funk}, {F{\"u}{\ss}ling}, {Gallant},
  {Giebels}, {Glicenstein}, {Gl{\"u}ck}, {Goret}, {Hadjichristidis}, {Hauser},
  {Hauser}, {Heinzelmann}, {Henri}, {Hermann}, {Hinton}, {Hoffmann}, {Hofmann},
  {Holleran}, {Hoppe}, {Horns}, {Jacholkowska}, {de Jager}, {Kendziorra},
  {Kerschhaggl}, {Kh{\'e}lifi}, {Komin}, {Kosack}, {Lamanna}, {Latham}, {Le
  Gallou}, {Lemi{\`e}re}, {Lemoine-Goumard}, {Lohse}, {Martin},
  {Martineau-Huynh}, {Marcowith}, {Masterson}, {Maurin}, {McComb}, {Moulin},
  {de Naurois}, {Nedbal}, {Nolan}, {Noutsos}, {Olive}, {Orford}, {Osborne},
  {Panter}, {Pelletier}, {Petrucci}, {Pita}, {P{\"u}hlhofer}, {Punch},
  {Ranchon}, {Raubenheimer}, {Raue}, {Rayner}, {Ripken}, {Rob}, {Rolland},
  {Rosier-Lees}, {Rowell}, {Sahakian}, {Santangelo}, {Saug{\'e}}, {Schlenker},
  {Schlickeiser}, {Schr{\"o}der}, {Schwanke}, {Schwarzburg}, {Schwemmer},
  {Shalchi}, {Sol}, {Spangler}, {Spanier}, {Steenkamp}, {Stegmann}, {Superina},
  {Tam}, {Tavernet}, {Terrier}, {Tluczykont}, {van Eldik}, {Vasileiadis},
  {Venter}, {Vialle}, {Vincent}, {V{\"o}lk}, {Wagner}, \&
  {Ward}}]{2007A&A...470..475A}
---. 2007, \aap, 470, 475

\bibitem[{{Albert} {et~al.}(2007){Albert}, {Aliu}, {Anderhub}, {Antoranz},
  {Armada}, {Baixeras}, {Barrio}, {Bartko}, {Bastieri}, {Becker}, {Bednarek},
  {Berger}, {Bigongiari}, {Biland}, {Bock}, {Bordas}, {Bosch-Ramon}, {Bretz},
  {Britvitch}, {Camara}, {Carmona}, {Chilingarian}, {Coarasa}, {Wibig},
  {Wittek}, {Zandanel}, {Zanin}, \& {Zapatero}}]{2007ApJ...666L..17A}
{Albert}, J., {et~al.} 2007, \apjl, 666, L17

\bibitem[{{Aliu} {et~al.}(2013){Aliu}, {Archambault}, {Arlen}, {Aune},
  {Beilicke}, {Benbow}, {Bird}, {Bouvier}, {Buckley}, {Bugaev}, {Cesarini},
  {Ciupik}, {Connolly}, {Cui}, {Dumm}, {Errando}, {Falcone}, {Federici},
  {Feng}, {Finley}, {Fortin}, {Fortson}, {Furniss}, {Galante}, {G{\'e}rard},
  {Gillanders}, {Griffin}, {Grube}, {Gyuk}, {Hanna}, {Holder}, {Hughes},
  {Humensky}, {Kaaret}, {Kertzman}, {Khassen}, {Kieda}, {Krawczynski},
  {Krennrich}, {Lang}, {Madhavan}, {Maier}, {Majumdar}, {McArthur}, {McCann},
  {Moriarty}, {Mukherjee}, {Nieto}, {O'Faol{\'a}in de Bhr{\'o}ithe}, {Ong},
  {Orr}, {Otte}, {Park}, {Perkins}, {Pohl}, {Popkow}, {Prokoph}, {Quinn},
  {Ragan}, {Reyes}, {Reynolds}, {Richards}, {Roache}, {Saxon}, {Sembroski},
  {Skole}, {Smith}, {Soares-Furtado}, {Staszak}, {Telezhinsky}, {Te{\v
  s}i{\'c}}, {Theiling}, {Varlotta}, {Vassiliev}, {Vincent}, {Wakely},
  {Weekes}, {Weinstein}, {Welsing}, {Williams}, {Zitzer}, {VERITAS
  Collaboration}, {B{\"o}ttcher}, {Fumagalli}, \&
  {Jadhav}}]{2013ApJ...779...92A}
{Aliu}, E., {et~al.} 2013, \apj, 779, 92

\bibitem[{{Aliu} {et~al.}(2014{\natexlab{a}}){Aliu}, {Archambault}, {Arlen},
  {Aune}, {Behera}, {Beilicke}, {Benbow}, {Berger}, {Bird}, {Bouvier},
  {Buckley}, {Bugaev}, {Byrum}, {Cerruti}, {Chen}, {Ciupik}, {Connolly}, {Cui},
  {Duke}, {Dumm}, {Errando}, {Falcone}, {Federici}, {Feng}, {Finley},
  {Fleischhack}, {Fortin}, {Fortson}, {Furniss}, {Galante}, {Gillanders},
  {Griffin}, {Griffiths}, {Grube}, {Gyuk}, {Hanna}, {Holder}, {Hughes},
  {Humensky}, {Johnson}, {Kaaret}, {Kertzman}, {Khassen}, {Kieda},
  {Krawczynski}, {Krennrich}, {Lang}, {Madhavan}, {Maier}, {Majumdar},
  {McArthur}, {McCann}, {Meagher}, {Millis}, {Moriarty}, {Mukherjee}, {Nieto},
  {O'Faol{\'a}in de Bhr{\'o}ithe}, {Ong}, {Otte}, {Park}, {Perkins}, {Pohl},
  {Popkow}, {Prokoph}, {Quinn}, {Ragan}, {Reyes}, {Reynolds}, {Richards},
  {Roache}, {Sembroski}, {Smith}, {Staszak}, {Telezhinsky}, {Theiling},
  {Varlotta}, {Vassiliev}, {Vincent}, {Wakely}, {Weekes}, {Weinstein},
  {Welsing}, {Williams}, {Zajczyk}, \& {Zitzer}}]{2014ApJ...782...13A}
---. 2014{\natexlab{a}}, \apj, 782, 13

\bibitem[{{Aliu} {et~al.}(2014{\natexlab{b}}){Aliu}, {Archambault}, {Arlen},
  {Aune}, {Barnacka}, {Beilicke}, {Benbow}, {Berger}, {Bird}, {Bouvier},
  {Buckley}, {Bugaev}, {Cerruti}, {Chen}, {Ciupik}, {Collins-Hughes},
  {Connolly}, {Cui}, {Dumm}, {Eisch}, {Falcone}, {Federici}, {Feng}, {Finley},
  {Fleischhack}, {Fortin}, {Fortson}, {Furniss}, {Galante}, {Gillanders},
  {Griffin}, {Griffiths}, {Grube}, {Gyuk}, {H{\aa}kansson}, {Hanna}, {Holder},
  {Hughes}, {Hughes}, {Humensky}, {Johnson}, {Kaaret}, {Kar}, {Kertzman},
  {Khassen}, {Kieda}, {Krawczynski}, {Krennrich}, {Lang}, {Madhavan},
  {Majumdar}, {McArthur}, {McCann}, {Meagher}, {Millis}, {Moriarty},
  {Mukherjee}, {Nelson}, {Nieto}, {O'Faol{\'a}in de Bhr{\'o}ithe}, {Ong},
  {Otte}, {Park}, {Perkins}, {Pohl}, {Popkow}, {Prokoph}, {Quinn}, {Ragan},
  {Rajotte}, {Reyes}, {Reynolds}, {Richards}, {Roache}, {Sadun}, {Santander},
  {Sembroski}, {Shahinyan}, {Sheidaei}, {Smith}, {Staszak}, {Telezhinsky},
  {Theiling}, {Tyler}, {Varlotta}, {Vassiliev}, {Vincent}, {Wakely}, {Weekes},
  {Weinstein}, {Welsing}, {Wilhelm}, {Williams}, {Zitzer}, {VERITAS
  Collaboration}, {B{\"o}ttcher}, \& {Fumagalli}}]{2014ApJ...797...89A}
---. 2014{\natexlab{b}}, \apj, 797, 89

\bibitem[{{Anderhub} {et~al.}(2009){Anderhub}, {Antonelli}, {Antoranz},
  {Backes}, {Baixeras}, {Balestra}, {Barrio}, {Bastieri}, {Becerra
  Gonz{\'a}lez}, {Becker}, {Sitarek}, {Sobczynska}, {Spanier}, {Spiro},
  {Stamerra}, {Stark}, {Suric}, {Takalo}, {Tavecchio}, {Temnikov}, {Tescaro},
  {Teshima}, {Torres}, {Turini}, {Vankov}, {Wagner}, {Villforth}, {Zabalza},
  {Zandanel}, {Zanin}, \& {Zapatero}}]{2009ApJ...704L.129A}
{Anderhub}, H., {et~al.} 2009, \apjl, 704, L129

\bibitem[{{Arnaud}(1996)}]{1996ASPC..101...17A}
{Arnaud}, K.~A. 1996, in Astronomical Society of the Pacific Conference Series,
  Vol. 101, Astronomical Data Analysis Software and Systems V, ed.
  {G.~H.~Jacoby \& J.~Barnes}, 17--+

\bibitem[{{Atoyan} \& {Dermer}(2004)}]{2004ApJ...613..151A}
{Atoyan}, A., \& {Dermer}, C.~D. 2004, \apj, 613, 151

\bibitem[{{Band} \& {Grindlay}(1985)}]{THEO::SSC_BAND}
{Band}, D.~L., \& {Grindlay}, J.~E. 1985, \apj, 298, 128

\bibitem[{{Bessell}(1990)}]{Bessel90}
{Bessell}, M.~S. 1990, \pasp, 102, 1181

\bibitem[{{B{\l}a{\.z}ejowski} {et~al.}(2000){B{\l}a{\.z}ejowski}, {Sikora},
  {Moderski}, \& {Madejski}}]{2000ApJ...545..107B}
{B{\l}a{\.z}ejowski}, M., {Sikora}, M., {Moderski}, R., \& {Madejski}, G.~M.
  2000, \apj, 545, 107

\bibitem[{Bonning {et~al.}(2012)Bonning, Urry, Bailyn, Buxton, Chatterjee,
  Coppi, Fossati, Isler, \& Maraschi}]{Smarts}
Bonning, E., {et~al.} 2012, The Astrophysical Journal, 756, 13

\bibitem[{{B{\"o}ttcher} {et~al.}(2008){B{\"o}ttcher}, {Dermer}, \&
  {Finke}}]{boett08}
{B{\"o}ttcher}, M., {Dermer}, C.~D., \& {Finke}, J.~D. 2008, \apjl, 679, L9

\bibitem[{{Breeveld} {et~al.}(2011){Breeveld}, {Landsman}, {Holland}, {Roming},
  {Kuin}, \& {Page}}]{2011AIPC.1358..373B}
{Breeveld}, A.~A., {Landsman}, W., {Holland}, S.~T., {Roming}, P., {Kuin},
  N.~P.~M., \& {Page}, M.~J. 2011, in American Institute of Physics Conference
  Series, Vol. 1358, American Institute of Physics Conference Series, ed. J.~E.
  {McEnery}, J.~L. {Racusin}, \& N.~{Gehrels}, 373--376

\bibitem[{{Burrows} {et~al.}(2005){Burrows}, {Hill}, {Nousek}, {Kennea},
  {Wells}, {Osborne}, {Abbey}, {Beardmore}, {Mukerjee}, {Short}, {Chincarini},
  {Campana}, {Citterio}, {Moretti}, {Pagani}, {Tagliaferri}, {Giommi},
  {Capalbi}, {Tamburelli}, {Angelini}, {Cusumano}, {Br{\"a}uninger}, {Burkert},
  \& {Hartner}}]{2005SSRv..120..165B}
{Burrows}, D.~N., {et~al.} 2005, \ssr, 120, 165

\bibitem[{{Cardelli} {et~al.}(1989){Cardelli}, {Clayton}, \&
  {Mathis}}]{1989ApJ...345..245C}
{Cardelli}, J.~A., {Clayton}, G.~C., \& {Mathis}, J.~S. 1989, \apj, 345, 245

\bibitem[{{Celotti} {et~al.}(2001){Celotti}, {Ghisellini}, \&
  {Chiaberge}}]{2001MNRAS.321L...1C}
{Celotti}, A., {Ghisellini}, G., \& {Chiaberge}, M. 2001, \mnras, 321, L1

\bibitem[{{Davis}(2001)}]{Davis2001}
{Davis}, J.~E. 2001, \apj, 562, 575

\bibitem[{{Falomo} {et~al.}(1993){Falomo}, {Bersanelli}, {Bouchet}, \&
  {Tanzi}}]{1993AJ....106...11F}
{Falomo}, R., {Bersanelli}, M., {Bouchet}, P., \& {Tanzi}, E.~G. 1993, \aj,
  106, 11

\bibitem[{{Finke} {et~al.}(2008){Finke}, {Dermer}, \&
  {B{\"o}ttcher}}]{2008ApJ...686..181F}
{Finke}, J.~D., {Dermer}, C.~D., \& {B{\"o}ttcher}, M. 2008, \apj, 686, 181

\bibitem[{{Franceschini} {et~al.}(2008){Franceschini}, {Rodighiero}, \&
  {Vaccari}}]{2008A&A...487..837F}
{Franceschini}, A., {Rodighiero}, G., \& {Vaccari}, M. 2008, \aap, 487, 837

\bibitem[{{Godet} {et~al.}(2009){Godet}, {Beardmore}, {Abbey}, {Osborne},
  {Cusumano}, {Pagani}, {Capalbi}, {Perri}, {Page}, {Burrows}, {Campana},
  {Hill}, {Kennea}, \& {Moretti}}]{2009A&A...494..775G}
{Godet}, O., {et~al.} 2009, \aap, 494, 775

\bibitem[{{Hardcastle}(2006)}]{2006MNRAS.366.1465H}
{Hardcastle}, M.~J. 2006, \mnras, 366, 1465

\bibitem[{{Harris} {et~al.}(2011){Harris}, {Massaro}, {Cheung}, {Horns},
  {Raue}, {Stawarz}, {Wagner}, {Colin}, {Mazin}, {Wagner}, {Beilicke},
  {LeBohec}, {Hui}, \& {Mukherjee}}]{2011ApJ...743..177H}
{Harris}, D.~E., {et~al.} 2011, \apj, 743, 177

\bibitem[{{Hervet} {et~al.}(2015){Hervet}, {Boisson}, \&
  {Sol}}]{2015arXiv150301377H}
{Hervet}, O., {Boisson}, C., \& {Sol}, H. 2015, ArXiv e-prints

\bibitem[{{Hofmann}(2010)}]{2010ATel.2743....1H}
{Hofmann}, W. 2010, The Astronomer's Telegram, 2743, 1

\bibitem[{{Hyv{\"o}nen} {et~al.}(2007){Hyv{\"o}nen}, {Kotilainen}, {Falomo},
  {{\"O}rndahl}, \& {Pursimo}}]{2007A&A...476..723H}
{Hyv{\"o}nen}, T., {Kotilainen}, J.~K., {Falomo}, R., {{\"O}rndahl}, E., \&
  {Pursimo}, T. 2007, \aap, 476, 723

\bibitem[{{Jahoda} {et~al.}(1996){Jahoda}, {Swank}, {Giles}, {Stark},
  {Strohmayer}, {Zhang}, \& {Morgan}}]{1996SPIE.2808...59J}
{Jahoda}, K., {Swank}, J.~H., {Giles}, A.~B., {Stark}, M.~J., {Strohmayer}, T.,
  {Zhang}, W., \& {Morgan}, E.~H. 1996, in Society of Photo-Optical
  Instrumentation Engineers (SPIE) Conference Series, Vol. 2808, Society of
  Photo-Optical Instrumentation Engineers (SPIE) Conference Series, ed.
  {O.~H.~Siegmund \& M.~A.~Gummin}, 59--70

\bibitem[{{Jones} {et~al.}(2009){Jones}, {Read}, {Saunders}, {Colless},
  {Jarrett}, {Parker}, {Fairall}, {Mauch}, {Sadler}, {Watson}, {Burton},
  {Campbell}, {Cass}, {Croom}, {Dawe}, {Fiegert}, {Frankcombe}, {Hartley},
  {Huchra}, {James}, {Kirby}, {Lahav}, {Lucey}, {Mamon}, {Moore}, {Peterson},
  {Prior}, {Proust}, {Russell}, {Safouris}, {Wakamatsu}, {Westra}, \&
  {Williams}}]{2009MNRAS.399..683J}
{Jones}, D.~H., {et~al.} 2009, \mnras, 399, 683

\bibitem[{{Jones}(1968)}]{JONES_KERNEL}
{Jones}, F.~C. 1968, Physical Review, 167, 1159

\bibitem[{{Kalberla} {et~al.}(2005){Kalberla}, {Burton}, {Hartmann}, {Arnal},
  {Bajaja}, {Morras}, \& {P{\"o}ppel}}]{2005A&A...440..775K}
{Kalberla}, P.~M.~W., {Burton}, W.~B., {Hartmann}, D., {Arnal}, E.~M.,
  {Bajaja}, E., {Morras}, R., \& {P{\"o}ppel}, W.~G.~L. 2005, \aap, 440, 775

\bibitem[Kaufmann et al.(2013)]{2013ApJ...776...68K} Kaufmann, S., Wagner, 
S.~J., \& Tibolla, O.\ 2013, \apj, 776, 68 


\bibitem[{{Lister} {et~al.}(1998){Lister}, {Marscher}, \&
  {Gear}}]{1998ApJ...504..702L}
{Lister}, M.~L., {Marscher}, A.~P., \& {Gear}, W.~K. 1998, \apj, 504, 702

\bibitem[{{Lister} {et~al.}(2009){Lister}, {Aller}, {Aller}, {Cohen}, {Homan},
  {Kadler}, {Kellermann}, {Kovalev}, {Ros}, {Savolainen}, {Zensus}, \&
  {Vermeulen}}]{2009AJ....137.3718L}
{Lister}, M.~L., {et~al.} 2009, \aj, 137, 3718

\bibitem[{{Lister} {et~al.}(2013){Lister}, {Aller}, {Aller}, {Homan},
  {Kellermann}, {Kovalev}, {Pushkarev}, {Richards}, {Ros}, \&
  {Savolainen}}]{2013AJ....146..120L}
---. 2013, \aj, 146, 120

\bibitem[{{Meyer} \& {Georganopoulos}(2014)}]{2014ApJ...780L..27M}
{Meyer}, E.~T., \& {Georganopoulos}, M. 2014, \apjl, 780, L27

\bibitem[{{Nolan} {et~al.}(2012){Nolan}, {Abdo}, {Ackermann}, {Ajello},
  {Allafort}, {Antolini}, {Atwood}, {Axelsson}, {Baldini}, {Ballet}, \&
  et~al.}]{2012ApJS..199...31N}
{Nolan}, P.~L., {et~al.} 2012, \apjs, 199, 31

\bibitem[{{Padovani} \& {Giommi}(1995)}]{1995ApJ...444..567P}
{Padovani}, P., \& {Giommi}, P. 1995, \apj, 444, 567

\bibitem[{{Paggi} {et~al.}(2009){Paggi}, {Massaro}, {Vittorini}, {Cavaliere},
  {D'Ammando}, {Vagnetti}, \& {Tavani}}]{2009A&A...504..821P}
{Paggi}, A., {Massaro}, F., {Vittorini}, V., {Cavaliere}, A., {D'Ammando}, F.,
  {Vagnetti}, F., \& {Tavani}, M. 2009, \aap, 504, 821

\bibitem[{{Planck Collaboration} {et~al.}(2011){Planck Collaboration}, {Ade},
  {Aghanim}, {Arnaud}, {Ashdown}, {Aumont}, {Baccigalupi}, {Balbi}, {Banday},
  {Barreiro}, \& et~al.}]{2011A&A...536A...7P}
{Planck Collaboration} {et~al.} 2011, \aap, 536, A7

\bibitem[{{Poole} {et~al.}(2008){Poole}, {Breeveld}, {Page}, {Landsman},
  {Holland}, {Roming}, {Kuin}, {Brown}, {Gronwall}, {Hunsberger}, {Koch},
  {Mason}, {Schady}, {vanden Berk}, {Blustin}, {Boyd}, {Broos}, {Carter},
  {Chester}, {Cucchiara}, {Hancock}, {Huckle}, {Immler}, {Ivanushkina},
  {Kennedy}, {Marshall}, {Morgan}, {Pandey}, {de Pasquale}, {Smith}, \&
  {Still}}]{2008MNRAS.383..627P}
{Poole}, T.~S., {et~al.} 2008, \mnras, 383, 627

\bibitem[{{Ravasio} {et~al.}(2002){Ravasio}, {Tagliaferri}, {Ghisellini},
  {Giommi}, {Nesci}, {Massaro}, {Chiappetti}, {Celotti}, {Costamante},
  {Maraschi}, {Tavecchio}, {Tosti}, {Treves}, {Wolter}, {Balonek}, {Carini},
  {Kato}, {Kurtanidze}, {Montagni}, {Nikolashvili}, {Noble}, {Nucciarelli},
  {Raiteri}, {Sclavi}, {Uemura}, \& {Villata}}]{2002A&A...383..763R}
{Ravasio}, M., {et~al.} 2002, \aap, 383, 763

\bibitem[{{Roming} {et~al.}(2009){Roming}, {Koch}, {Oates}, {Porterfield},
  {Vanden Berk}, {Boyd}, {Holland}, {Hoversten}, {Immler}, {Marshall}, {Page},
  {Racusin}, {Schneider}, {Breeveld}, {Brown}, {Chester}, {Cucchiara},
  {DePasquale}, {Gronwall}, {Hunsberger}, {Kuin}, {Landsman}, {Schady}, \&
  {Still}}]{2009ApJ...690..163R}
{Roming}, P.~W.~A., {et~al.} 2009, \apj, 690, 163

\bibitem[{{Sambruna} {et~al.}(2004){Sambruna}, {Gambill}, {Maraschi},
  {Tavecchio}, {Cerutti}, {Cheung}, {Urry}, \& {Chartas}}]{2004ApJ...608..698S}
{Sambruna}, R.~M., {Gambill}, J.~K., {Maraschi}, L., {Tavecchio}, F.,
  {Cerutti}, R., {Cheung}, C.~C., {Urry}, C.~M., \& {Chartas}, G. 2004, \apj,
  608, 698

\bibitem[{{Sanchez} {et~al.}(2013){Sanchez}, {Fegan}, \&
  {Giebels}}]{2013arXiv1303.5923S}
{Sanchez}, D.~A., {Fegan}, S., \& {Giebels}, B. 2013, ArXiv e-prints

\bibitem[{{Schlafly} \& {Finkbeiner}(2011)}]{2011ApJ...737..103S}
{Schlafly}, E.~F., \& {Finkbeiner}, D.~P. 2011, \apj, 737, 103

\bibitem[{Shaw {et~al.}(2013)Shaw, Romani, Cotter, Healey, Michelson, Readhead,
  Richards, Max-Moerbeck, King, \& Potter}]{0004-637X-764-2-135}
Shaw, M.~S., {et~al.} 2013, The Astrophysical Journal, 764, 135

\bibitem[{{Sikora} {et~al.}(1994){Sikora}, {Begelman}, \&
  {Rees}}]{1994ApJ...421..153S}
{Sikora}, M., {Begelman}, M.~C., \& {Rees}, M.~J. 1994, \apj, 421, 153

\bibitem[{{Silva} {et~al.}(1998){Silva}, {Granato}, {Bressan}, \&
  {Danese}}]{1998ApJ...509..103S}
{Silva}, L., {Granato}, G.~L., {Bressan}, A., \& {Danese}, L. 1998, \apj, 509,
  103

\bibitem[{{Tavecchio} {et~al.}(2010){Tavecchio}, {Ghisellini}, {Ghirlanda},
  {Foschini}, \& {Maraschi}}]{2010MNRAS.401.1570T}
{Tavecchio}, F., {Ghisellini}, G., {Ghirlanda}, G., {Foschini}, L., \&
  {Maraschi}, L. 2010, \mnras, 401, 1570

\bibitem[{{Tavecchio} {et~al.}(1998){Tavecchio}, {Maraschi}, \&
  {Ghisellini}}]{1998ApJ...509..608T}
{Tavecchio}, F., {Maraschi}, L., \& {Ghisellini}, G. 1998, \apj, 509, 608

\bibitem[{{Tavecchio} {et~al.}(2000){Tavecchio}, {Maraschi}, {Sambruna}, \&
  {Urry}}]{2000ApJ...544L..23T}
{Tavecchio}, F., {Maraschi}, L., {Sambruna}, R.~M., \& {Urry}, C.~M. 2000,
  \apjl, 544, L23

\bibitem[{{Tavecchio} {et~al.}(2007){Tavecchio}, {Maraschi}, {Wolter},
  {Cheung}, {Sambruna}, \& {Urry}}]{2007ApJ...662..900T}
{Tavecchio}, F., {Maraschi}, L., {Wolter}, A., {Cheung}, C.~C., {Sambruna},
  R.~M., \& {Urry}, C.~M. 2007, \apj, 662, 900

\bibitem[{{The Fermi-LAT Collaboration}(2013)}]{2013arXiv1306.6772T}
{The Fermi-LAT Collaboration}. 2013, ArXiv e-prints 1306.6772

\bibitem[{{Tsujimoto} {et~al.}(2011){Tsujimoto}, {Guainazzi}, {Plucinsky},
  {Beardmore}, {Ishida}, {Natalucci}, {Posson-Brown}, {Read}, {Saxton}, \&
  {Shaposhnikov}}]{2011A&A...525A..25T}
{Tsujimoto}, M., {et~al.} 2011, \aap, 525, A25

\bibitem[{{Vaughan} {et~al.}(2003){Vaughan}, {Edelson}, {Warwick}, \&
  {Uttley}}]{Vaughan}
{Vaughan}, S., {Edelson}, R., {Warwick}, R.~S., \& {Uttley}, P. 2003, \mnras,
  345, 1271

\bibitem[{{Wright}(2006)}]{2006PASP..118.1711W}
{Wright}, E.~L. 2006, \pasp, 118, 1711

\bibitem[{{Wright} {et~al.}(2010){Wright}, {Eisenhardt}, {Mainzer}, {Ressler},
  {Cutri}, {Jarrett}, {Kirkpatrick}, {Padgett}, {McMillan}, {Skrutskie},
  {Stanford}, {Cohen}, {Walker}, {Mather}, {Leisawitz}, {Gautier}, {McLean},
  {Benford}, {Lonsdale}, {Blain}, {Mendez}, {Irace}, {Duval}, {Liu}, {Royer},
  {Heinrichsen}, {Howard}, {Shannon}, {Kendall}, {Walsh}, {Larsen}, {Cardon},
  {Schick}, {Schwalm}, {Abid}, {Fabinsky}, {Naes}, \& {Tsai}}]{WISE}
{Wright}, E.~L., {et~al.} 2010, \aj, 140, 1868

\bibitem[{{Zensus} {et~al.}(2002){Zensus}, {Ros}, {Kellermann}, {Cohen},
  {Vermeulen}, \& {Kadler}}]{2002AJ....124..662Z}
{Zensus}, J.~A., {Ros}, E., {Kellermann}, K.~I., {Cohen}, M.~H., {Vermeulen},
  R.~C., \& {Kadler}, M. 2002, \aj, 124, 662

\end{thebibliography}
\bibliographystyle{mn2e}
\appendix
\section{Constraints for an arbitrary field of seed photons}
\label{appendix}

In leptonic class models, the inverse Compton process is responsible for the high energy part of the SED. The seed photons originate either from synchrotron
radiation produced within the jet (SSC models) or from a source outside of the jet (external Compton models). In the latter case, the sources can be either the broad-line regions or the dust torus \citep{1994ApJ...421..153S,2000ApJ...545..107B}.

The peak observed energy $E_{\rm S}$ of an electron with Lorentz factor $\gamma$ is given by $$E_{\rm s}/\mec=\frac{\delta\gamma^2B}{(1+z)B_{cr}},$$ and the Compton-scattered photon energy by $$E_{\rm ic}/\mec=\frac{\delta\gamma^2\epsilon_{\rm seed}'}{(1+z)},$$ where the energy\footnote{The notation $E = \epsilon\mec$ is adopted
  here.} of the seed photons is $\epsilon_{\rm seed} $ (respectively $\epsilon_{\rm seed}' = \delta \epsilon_{\rm seed}$ in the
jet's frame).

Efficient Compton scattering will occur only for electrons below the KN limit:

\begin{equation}
\gamma\leq(4\epsilon_{\rm seed}')^{-1}.
\label{eq:kn}
\end{equation}
This KN limit means that Compton-scattered photons will be mainly restricted to energies:
$$E_{\rm ic}/\mec\leq\frac{\delta}{16(1+z)\epsilon_{\rm seed}'}.$$
The synchrotron photons produced by the electrons having the energy $(4\epsilon_{\rm seed}')^{-1}$ have a peak energy given by:
$$E_{\rm s}/\mec=\frac{\delta B}{16(1+z)\epsilon_{\rm seed}'^2B_{cr}}.$$
Combining the last two equations with the constraints on maximal values for $E_{\rm s}\approx0.1$ keV and
$E_{\rm ic}\approx 1$ TeV derived from the observations yields:

\begin{equation}
\frac{\delta}{70}\geq\frac{B}{10^{-2} G},
\label{eq:res}
\end{equation}
which requires either an unusually high Doppler factor, or an unusually low magnetic field. If the 1 TeV photons are produced by IC scattering in the KN regime, Eq. \ref{eq:kn} becomes $$\gamma\geq(4\epsilon_{\rm seed}')^{-1}$$ and the observed photon energy is \citep{1998ApJ...509..608T} $$E_{\rm c}/\mec=\frac{\delta\gamma}{(1+z)}.$$ Then Eq. \ref{eq:res} reads 

\begin{equation}
\frac{\delta}{17.5}\leq\frac{B}{10^{-2} G}
\end{equation}
which is a reasonable constraint. Note that this calculation applies no matter what the seed photon source is
(broad-line region or dust torus or synchrotron photons produced within the jet), illustrating the
difficulties of either radiative scenarios to account for the main SED features of AP Librae in the Thomson
regime.

\section{Candidates for VHE observations}
\label{appendix2}

The detection of AP~Librae by the \hess\ telescopes has revealed the broadest IC component for a blazar with a peak position at very low energy. Unfortunately, only a handful of LBL-type objects have yet been detected at VHEs. To decide if AP~Librae is a special case or a typical representative of the LBL class, other LBL objects have to be observed by \v{C}erenkov telescope and detected at VHE.

Due to their limited field of view ($\approx5\deg$), an extra-galactic survey performed by \v{C}erenkov telescopes is not possible yet. As a consequence, good targets for observations have to be found based on multi-wavelengths data. In this appendix, six LBL-type objects, present in the second catalog of \fer\ sources \citep[2FGL,][]{2012ApJS..199...31N} were selected based on their possible VHE emission. The 2FGL best fit power-law, measured in the 100~MeV-100~GeV band, was extrapolated above 200~GeV and EBL correction was made based on the \citet{2008A&A...487..837F} model. The redshift information was extracted either from the second catalog of AGN \citep[2LAC,][]{2011ApJ...743..171A} or from \citet{0004-637X-764-2-135}. Sources without redshift measurement were excluded and only sources classified as a BL~Lac of the LBL class were retained. Note that AP~Librae appeared to be the first on this list when building it.

The names of six candidates, ranked by predicted flux above 200~GeV, are given in Table~\ref{table:prop}. For illustration, their SEDs, built from archival data using the ASDC SED builder\footnote{http://tools.asdc.asi.it/SED/}, are presented in Figure~\ref{Proposal}. Two out of the six sources can be observed by \hess\ and five by the northern facilities (VERITAS and MAGIC). Despite its location and with a redshift of $z=0.424$, the source 2FGL~J0738.0+1742 can be well suited for \hess\ II telescope observations given the lower energy threshold (50~GeV) of the instrument. The redshifts of 2FGL~J1150.1+2419 and 2FGL~J1150.1+2419, found in the 2LAC, were not confirmed by \citet{0004-637X-764-2-135}. Five out of 6 are also present in the first Fermi-LAT Catalog of Sources Above 10 GeV \citep[1FHL,][ see Figure~\ref{Proposal}]{2013arXiv1306.6772T}.

\begin{table*}
 \centering
 \begin{minipage}{150mm}
\caption{Proposed LBL-type objects for VHE observations. The 2FGL name is given in the first column with the position in the second and third. The redshift measurement taken from \citep{2011ApJ...743..171A} or \citep{0004-637X-764-2-135} is reported in the fourth column. The name (column 5) of the counterpart associated with the 2FGL source was found in the 2LAC catalog. The last column in the mane of the best suited instrument for observations. The sources are ranked by predicted flux above 200~GeV.} 
\label{table:prop}
  \begin{tabular}{@{}l c c c c c@{}}
  \hline

2FGL name&$\alpha_{\rm J2000}$&$\delta_{\rm J2000}$&redshift&Association&Instruments\\
  \hline
2FGL~J1719.3+1744 & \HMS{17}{19}{13.05} & \DMS{17}{45}{06.4}& 0.137& PKS 1717+177 & VERITAS/MAGIC \\
2FGL~J0617.6-1716& \HMS{06}{17}{ 33.67}	& \DMS{-17}{ 15}{ 22.8}& 0.098& CRATES J061733.67-171522.8 & \hess \\
2FGL~J0738.0+1742 & \HMS{ 07}{ 38}{ 07.39}& \DMS{17}{42}{19.0} & 0.424 & PKS 0735+17& VERITAS/MAGIC - \hess\\
2FGL~J1559.0+5627 & \HMS{15 }{58 }{48.29}& \DMS{56}{25}{14.1} & 0.3 & TXS 1557+565& VERITAS/MAGIC \\
2FGL~J1150.1+2419 & \HMS{ 11}{ 50 }{19.21}& \DMS{24}{17}{53.8} &0.2 & B2 1147+24& VERITAS/MAGIC \\
2FGL~J0712.9+5032 & \HMS{07}{ 12}{ 43.68}& \DMS{ 50}{ 33 }{22.7}& 0.502 &GB6 J0712+5033& VERITAS/MAGIC \\

\hline
\end{tabular}
\end{minipage}
\end{table*}

\newlength{\specwidth}
\setlength{\specwidth}{0.49\textwidth}
\newcommand{\includeSpec}[1]{\includegraphics[bb=40 2 550 400,clip,width=\specwidth]{#1}}

\begin{figure*}
\centering
\includeSpec{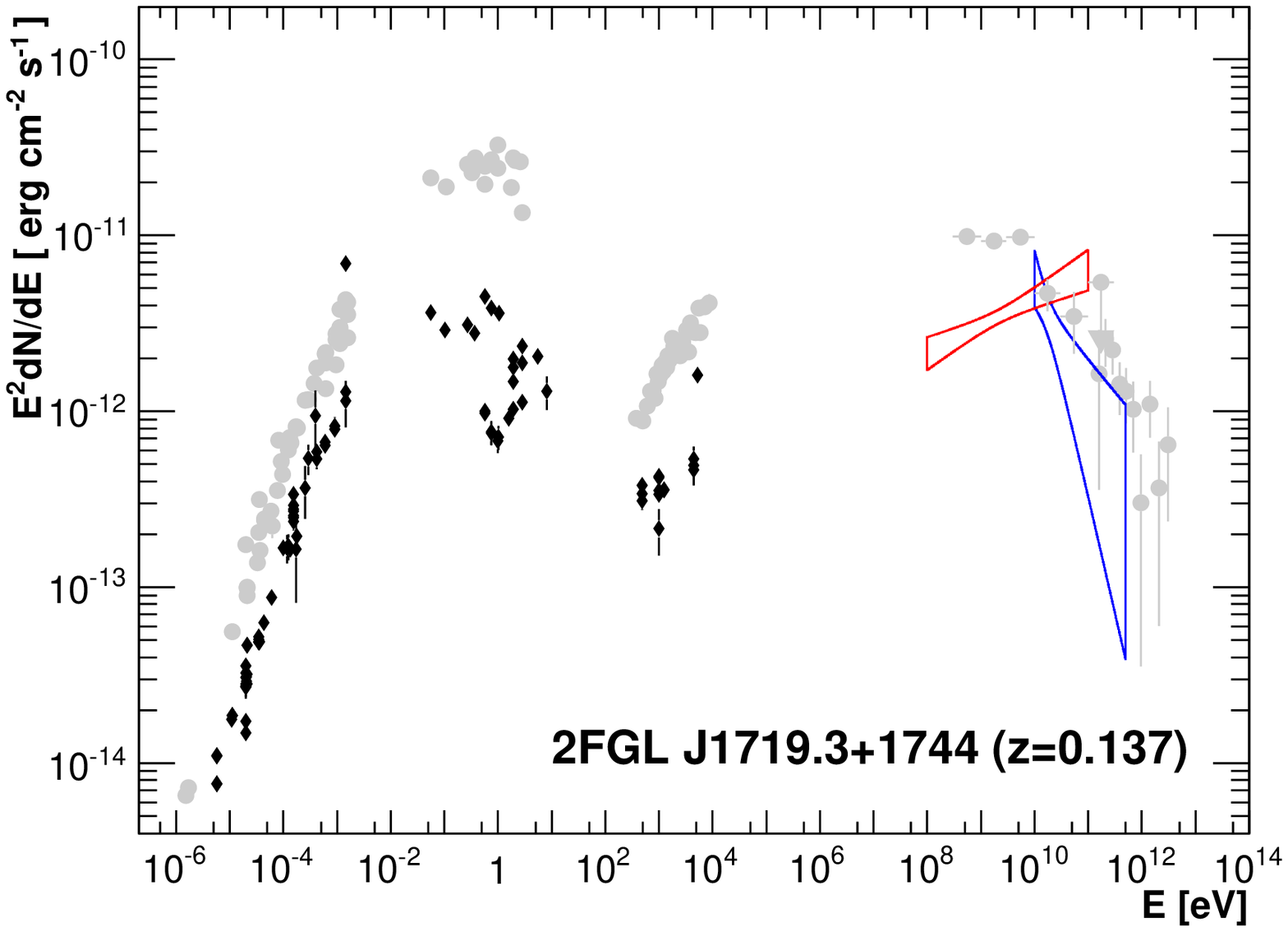}%
\includeSpec{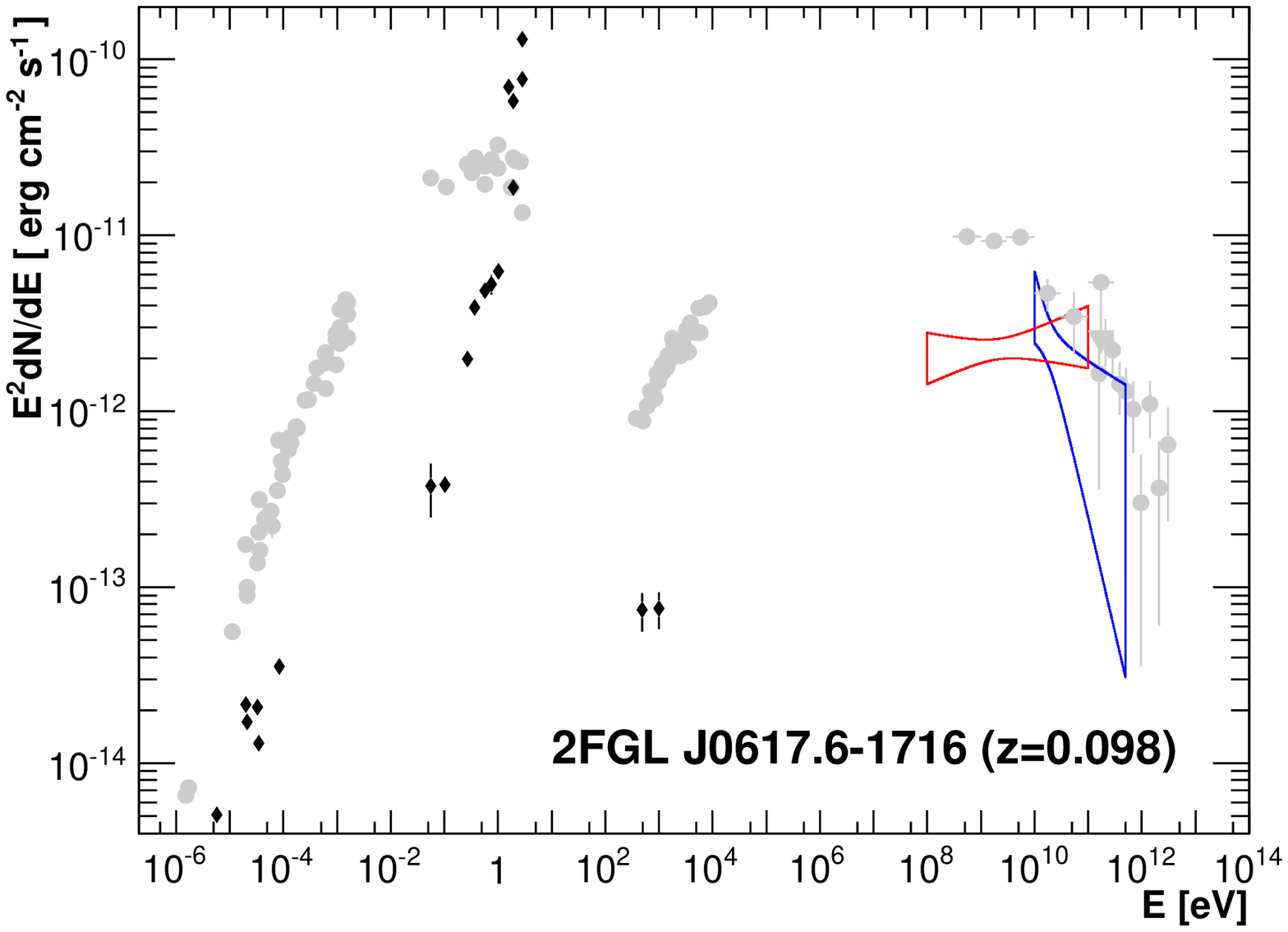}%

\includeSpec{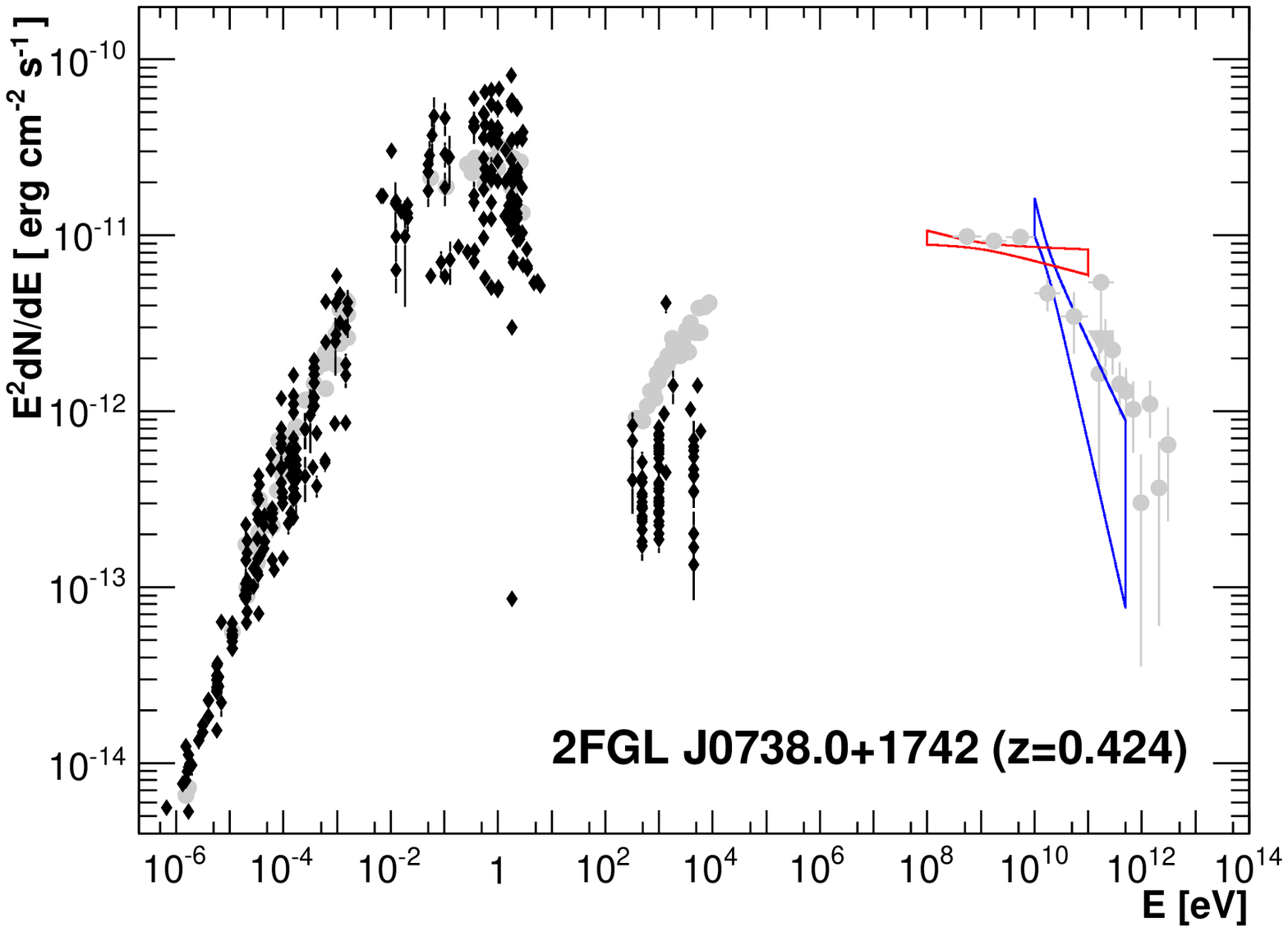}%
\includeSpec{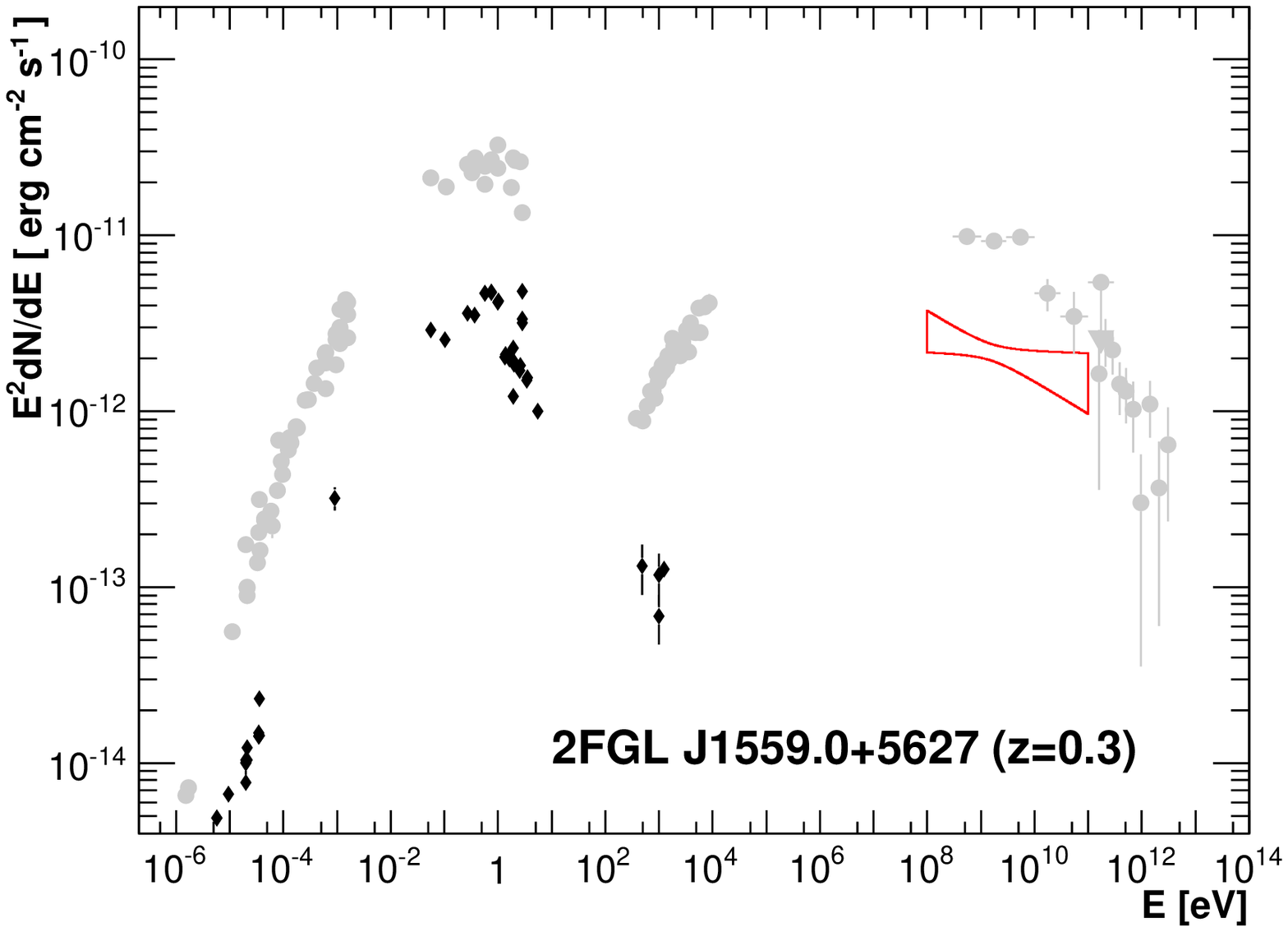}%

\includeSpec{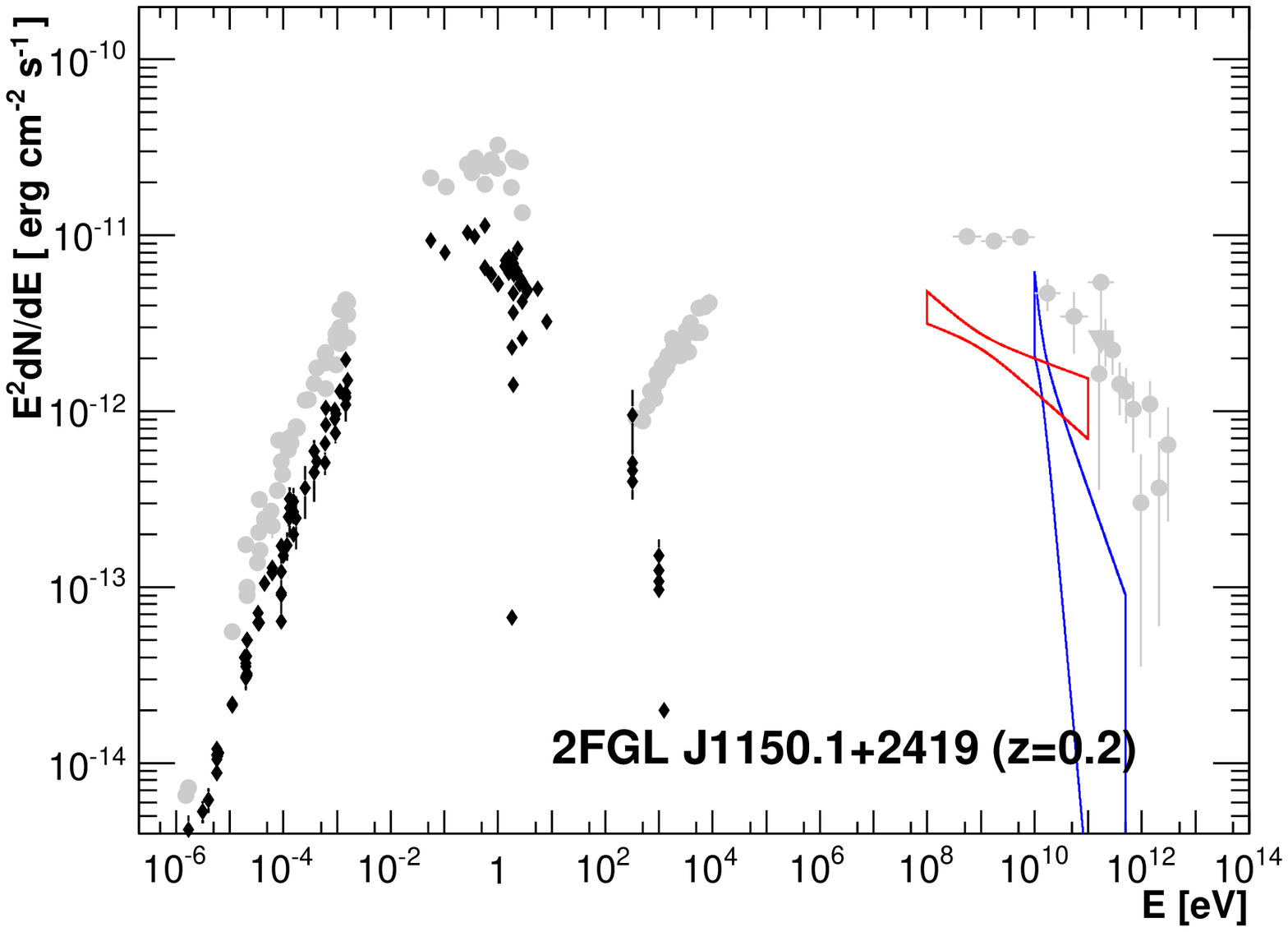}%
\includeSpec{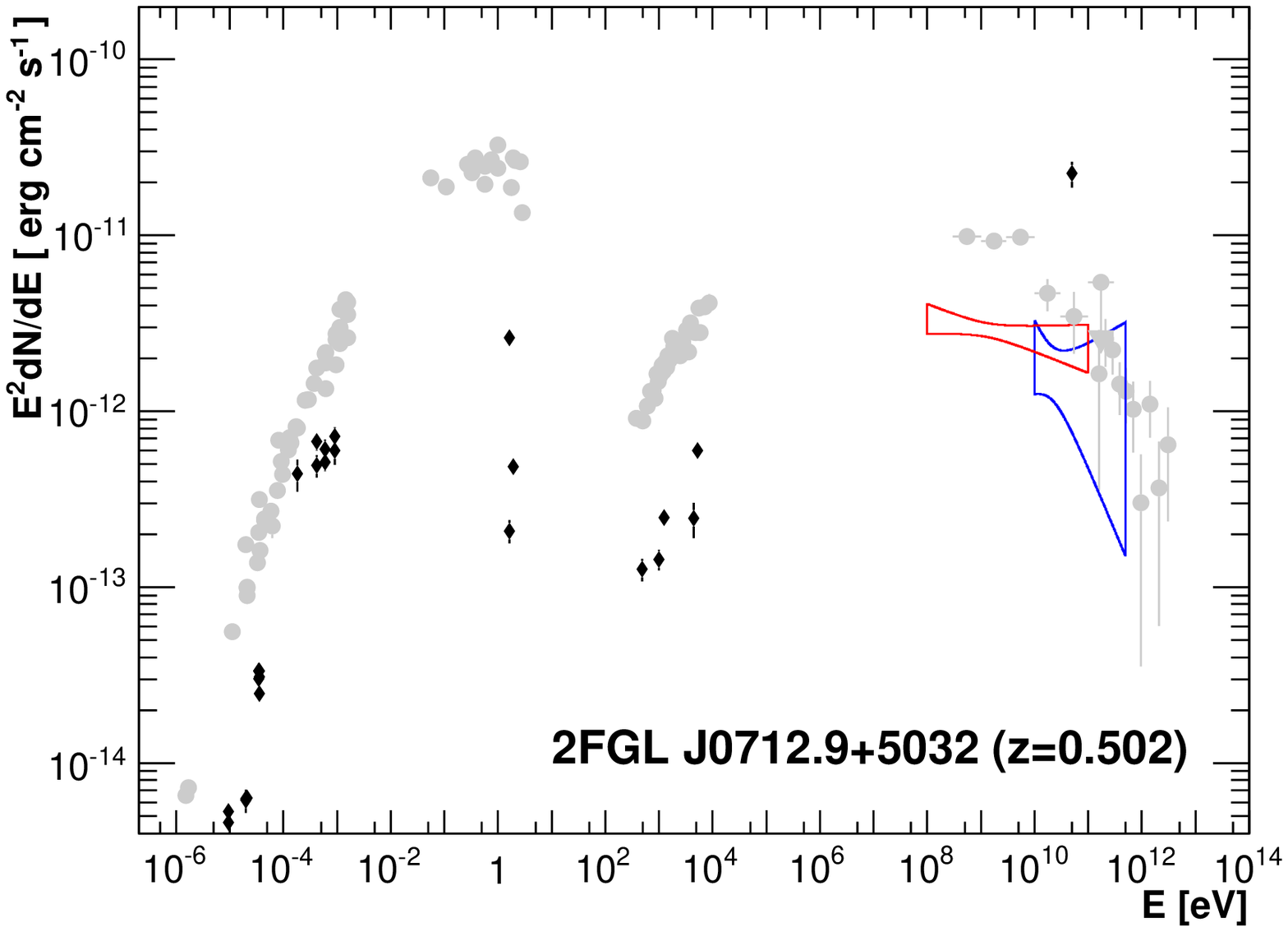}%

\caption{SEDs for the six LBL objects selected. The black points are archival data while the respectively red and blue butterflies are the 2FGL and 1FHL measurements. Gray points are the AP Librae data presented in this work.\label{Proposal}}
\end{figure*}

\end{document}